\documentclass[12pt,preprint]{aastex}

\def\fp1{\mbox{${\rm Log}(R_e)= a\ {\rm Log}(\sigma) + b\ \langle \mu \rangle_e +c$}}
\def\muem{\mbox{$\langle \mu \rangle_{\rm e}$}}
\def\re{\mbox{$R_{\rm e}$}}
\def\muere{\mbox{$\langle\mu\rangle_{\rm e} - {\rm Log}(R_{\rm e})$}}  
\def\H0{\mbox{$H_0$}}
\def\q0{\mbox{$q_0$}}
\def\Ob{\mbox{$\Omega_b$}}
\def\Ol{\mbox{$\Omega_\lambda$}}
\def\rq{\mbox{$r^{1/4}$}}
\def\kms{\rm km~s$^{-1}$}
\def\etal{et al.\/}
\def\eg{{\it e.g.\/}}

\shorttitle{The FP of the WINGS clusters}
\shortauthors{D'Onofrio et al.}

\begin{document}

\title{THE FUNDAMENTAL PLANE OF EARLY--TYPE GALAXIES IN NEARBY CLUSTERS FROM THE WINGS DATABASE}

\author{M. D'Onofrio\altaffilmark{1}, G. Fasano\altaffilmark{2}, J. Varela\altaffilmark{3}, D. Bettoni\altaffilmark{2}, M. Moles\altaffilmark{3}, P. Kj{\ae}rgaard\altaffilmark{4}, E. Pignatelli\altaffilmark{5}, B. Poggianti\altaffilmark{2}, A. Dressler\altaffilmark{6}, A. Cava\altaffilmark{2}, J. Fritz\altaffilmark{2}, W.J. Couch\altaffilmark{7}, A. Omizzolo\altaffilmark{2,8}}

\affil{$^1$Astronomy Department, Vicolo Osservatorio 3, I-35122 Padova, Italy}
\affil{$^2$INAF/Astronomical Observatory of Padova, Vicolo Osservatorio 2, I-35122 Padova, Italy}
\affil{$^3$Instituto de Astrofisica de Andalusia (C.S.I.C.), Apartado 2004, 1 8080 Granada, Spain}
\affil{$^4$Copenhagen University Observatory. The Niels Bohr Institute for Astronomy Physics and Geophysics, Juliane Maries Vej 30, 2100, Copenhagen, Denmark}
\affil{$^5$Liceo scientifico statale I. Nievo, via Barbarigo 38, Padova, Italy}
\affil{$^6$Observatories of the Carnegie Institution of Washington, Pasadena, CA 91101, USA}
\affil{$^7$School of Physics, University of New South Wales, Sydney 2052, Australia}
\affil{$^8$Vatican Observatory Research Group, University of Arizona, Tucson, AZ 85721, USA }

\email{mauro.donofrio@unipd.it}

\begin{abstract}

By exploiting the data of three large surveys (WINGS, NFPS and SDSS),
we present here a comparative analysis of the Fundamental Plane \fp1
(FP hereafter) of the early-type galaxies (ETGs) belonging to 59
galaxy clusters in the redshift range $0.04<z<0.07$.

We show that the variances of the distributions of the FP coefficients
derived for the clusters in our sample are just marginally consistent
with the hypothesis of universality of the FP relation. By
investigating the origin of such remarkable variances we find that,
besides a couple of obvious factors, such as the adopted fitting
technique and the method used to measure the photometric and kinematic
variables, the coefficients of the FP are also influenced by the
distribution of photometric/kinematic properties of galaxies in the
particular sample under analysis. In particular, some indication is
found that the FP coefficients intrinsically depend on the particular
luminosity range of the sample, suggesting that bright and faint ETGs
could have systematically different FPs. We speculate that the FP is
actually a bent surface, which is approximated by different planes
when different selection criteria, either chosen or induced by
observations, are acting to define galaxies samples.

We also find strong correlations between the FP coefficients and the
local cluster environment (cluster-centric distance and local
density), while the correlations with the galaxy properties are less
marked (Sersic index), weak (color) or even absent
(flattening). Furthermore, the FP coefficients appear to be poorly
correlated with the global properties of clusters, such as richness,
virial radius, velocity dispersion, optical and X-ray luminosity.

The relation between luminosity and mass of our galaxies, computed by
tacking into account the deviations from the \rq\ light profiles
(Sersic profiles), indicates that, for a given mass, the greater the
light concentration (high Sersic index $n$) the higher the luminosity,
while, for a given luminosity, the lower the light concentration, the
greater the mass. Moreover, the relation between mass-to-light ratio
and mass for our galaxy sample (with Sersic profile fitting) turns out
to be steeper and broader than that obtained for the Coma cluster
sample with \rq\ profile fitting. This broadness, together with the
bending we suspect to be present in the FP, might partly
reconcile the phenomenology of the scaling relations of ETGs with the
expectations from the $\Lambda$CDM cosmology.

The present analysis indicates that the claimed universality of the FP
of ETGs in clusters is still far from being proven and that systematic
biases might affect the conclusions found in the literature about the
luminosity evolution of ETGs, since datasets at different redshifts
and with different distributions of the photometric/kinematic
properties of galaxies are compared with each other.

\end{abstract}


\section{INTRODUCTION}\label{intro}

The survey WINGS \citep{Fas06} is providing a huge amount of
spectroscopic and photometric (multi-band) data for several thousands
galaxies in a complete sample of X-Ray selected clusters in the local
Universe (0.04$<$z$<$0.07). Among the other things, line indices and
equivalent widths (including Mg2 line-strengths) of galaxies are going
to be available for $\sim$6,000 galaxies, while, for $\sim$40,000
galaxies, we already have at our disposal the structural parameters
(\re\ , \muem\ and Sersic index $n$) derived using the automatic
surface photometry tool GASPHOT (\citealp{Pignatelli}). This put us in
a privileged position to analyse the scaling relations of nearby
cluster galaxies with unprecedented statistical robustness. In this
paper we will focus on the Fundamental Plane of early-type galaxies.

Since its discovery, the FP relation: \fp1 \citep{7Sam,DjDa} 
has been widely used as a tool to investigate the
properties of ETGs, to derive cluster distances and galaxy peculiar
motions (see \eg\ the ENACS cluster survey of \citealp{Katgert}, the
SMAC survey of \citealp{SMAC}, and the EFAR project of \citealp{Wegner}), to
perform cosmological tests and compute cosmological parameters (see
\eg\ \citealp{Moles98}), and as a diagnostic tool of galaxy evolution
and $M/L$ variations with redshift (see \eg\ \citealp{KJM} and
\citealp{Ziegl99}). Most analyses in the literature are based on the
comparison between the FP of distant clusters and that of nearby
clusters, usually set on the Coma cluster (\citealp{Jorg96}), the only
one with extensive, homogeneous photometric and spectroscopic data
for a large sample of ETGs.

Even if the universality of the FP relation at low redshift has never
been actually proven, it has been recently claimed that the FP
coefficient $a$\footnote{This coefficient is related to the tilt of
the FP, represented by the difference $2-a$, that is the deviation
from the Virial expectation value $a=2$.} changes systematically at
increasing redshift, from $\sim 1.2$ at redshift zero to $\sim0.8$ at
$z\sim0.8\div1.3$ \citep{Sperello,Jorg06}. This change, already
predicted by \citet{Pahrea}, has been attributed to the evolution of
ETGs with redshift.

However, the situation is far from being clear, since the data
required to assess the universality of the FP are still lacking. The
SDSS survey \citep{sloan} first attempted to face this problem
adopting the correct strategy, which must necessarily rest on the
availability of large galaxy samples. The results of this analysis
indicate that the FP is a robust relation valid for all ETGs (above
the magnitude limit of the SDSS), but its coefficients could depend on
the number density of the galaxy environment: the luminosities, sizes,
and velocity dispersions of the ETGs seem to increase slightly as the
local density increases, while the average surface brightnesses
decrease. However, evidences supporting different conclusions have
been found by \citet{delaRosa}, \citet{Pahrea,Pahreb} and
\citet{Kochanek}.

In addition, it is still unclear whether ETGs in clusters at the
same redshift share the same FP, or instead the FP coefficients
systematically change as a function of the global properties of the 
host cluster (richness, optical and X-ray luminosity, velocity
dispersions, concentration, subclustering, etc.).

Today, thanks to the huge observational effort done by wide field
surveys, such as SDSS \citep{sloan}, NFPS \citep{noao} and WINGS
\citep{Fas06}, the study of the FP can be extended to a much larger
sample of nearby clusters. Besides the data from the SDSS survey, we
can now use those from two more surveys (WINGS and NFPS) suitably
designed to study the properties of nearby clusters. Here we exploit
these datasets to check whether, at least in the local Universe, the
hypothesis of universality of the FP turns out to be supported by the
observations or not.

The paper is structured as follows: in Sec.~\ref{sec1} we present our
data sample, discussing its properties, its statistical completeness
and the intrinsic uncertainties associated to the measured structural
(effective radius), photometric (effective surface brightness) and
dynamical (central velocity dispersion) quantities involved in the FP
relation. In Sec.~\ref{sec2} we present the FP for the whole dataset
and those of each individual cluster. In Sec.~\ref{sec3}, also by
means of extensive simulations, we investigate the origin of the large
spread observed in the FP coefficients, showing that the scatter is
hardly attributable just to the statistical uncertainty arising
from the limited number of ETGs in each cluster. In Sec.~\ref{sec4} we
explore the behaviour of the FP coefficients at varying some galaxy
properties (Sersic index, color, flattening), the local environment
(cluster-centric distance and local galaxy density) and the global
properties of the host clusters (density, central velocity dispersion,
optical and X-ray luminosity). Finally, in Sec.~\ref{sec5}, we discuss
the relations involving the mass and the mass-to-light ratio of ETGs
in nearby clusters, which are closely linked to the FP, also providing
a tool to investigate the galaxy formation and evolution. Conclusions
are drawn in Sec.~\ref{sec6}.
Hereafter in this paper we adopt the standard cosmological parameters
$\H0=70$, $\Ol=0.7$, $\Ob=0.3$.

\section{THE GALAXY SAMPLE}\label{sec1}

\placetable{tbl1}

The initial galaxy sample has been extracted from 59 clusters
belonging to the survey WINGS~(W). It includes galaxies having
velocity dispersion measurements and 'early-type' classifications from
the surveys SDSS~(S) and/or NFPS~(N). Effective radius and
surface brightness of galaxies have been measured by GASPHOT
(\citealp{Pignatelli}), the software purposely devised to perform the
surface photometry of galaxies with threshold isophotal area (at
2$\times rms_{\rm bkg}$) larger than 200 pixels in the WINGS
survey (Pignatelli et al. in preparation). The central velocity
dispersions have been extracted from the catalogs published by the
surveys NFPS (52 clusters in common with WINGS) and SDSS (14 clusters
in common with WINGS). The clusters in common between NFPS, SDSS and
WINGS are: A0085, A119, A160, A602, A957x, A2124, and A2399.

A careful check of morphologies, performed both visually and using the
automatic tool MORPHOT (Fasano et al. in preparation; again purposely
devised for the WINGS survey), allowed us to identify in both datasets
several early-type spirals, erroneously classified as E or S0 galaxies
($\sim 8$\% of the whole sample). Besides these, we also decided to
exclude from the present analysis the galaxies with central velocity
dispersion $\sigma<95$ \kms (see Sec.\ref{sec12}) or total
luminosity $M_V>-18$. The final sample sizes are: $N_{W+N}$=1368;
$N_{W+S}$=282; $N_{W+N+S}$=1550 (100 objects in common between W+S and
W+N). The median number of ETGs per cluster is $N_{med}=23$.
For each cluster, Table~\ref{tbl1} reports the number of galaxies in 
the two samples (W+N and W+S; columns 8 and 9, respectively) and that 
of galaxies in common (W+[N\&S]; column 10).

The table also reports some salient cluster properties: average
redshift (column 2; from NED), velocity dispersion ($\Sigma$) of
galaxies around the average redshift (column 3; again from NED), X-ray
(0.1-2.4~keV) luminosity in ergs~s$^{-1}$ (column 4; from
\citealp{ebel1,ebel2,ebel3}), total absolute magnitude in the V-band
(column 5; from the WINGS deep catalogs), radius $R_{200}$ in Mpc
(column 6; from $\Sigma$, following \citealp{pogg06}) and absolute
V-band magnitude of the brightest cluster member (column 7; again from
the WINGS catalogs).

It is worth stressing that, even though our sample of ETGs is the most
sizeable among those used till now to study the FP of nearby clusters,
it is still far from being complete from a statistical point of
view. In particular: (i) the surface photometry is available just for
the galaxies in the region of $\sim35\times35$ arcminutes around the
cluster center (the regions mapped by the CCD images of the WINGS
survey); (ii) the SDSS and NFPS surveys have provided velocity
dispersions just for subsamples of the WINGS ETGs, each survey
according to the proper selection criteria (see Sec.\ref{sec12});
(iii) a couple of clusters with SDSS velocity dispersions are just
partially mapped by the survey.

\subsection{The WINGS photometry}\label{sec11}

The WINGS survey has produced catalogs of deep photometry and surface
photometry for 77 nearby clusters. For several thousands galaxies per
cluster the deep catalogs contain many geometrical and aperture
photometry data (Varela et al. 2008, A\&A, in press.), derived by means
of SExtractor analysis (\citealp{Bertin}). The surface photometry
catalogs contain data for several hundreds galaxies per cluster (those
with isophotal area greater than 200 pixels) and have been produced by
using the previously mentioned tool GASPHOT. For each galaxy it
performs seeing convolved, simultaneous, Sersic law fitting of the
major and minor axis growth profiles, thus providing Sersic index $n$,
effective radius \re\ and average surface brightness \muem\ , total
luminosity, flattening and local sky background. The data and the
associated uncertainties are discussed in Pignatelli et al. (2008, in
preparation). The average quoted $rms$ uncertainties of \re\ and
\muem\ are $\sim 15\%$ and $\sim 10\%$, respectively. The surface
brightnesses have been corrected for galactic extinction
(\citealp{Schlegel}) and cosmological dimming (using the average
redshifts of the clusters), while the K-corrections have not been
considered. Effective radii have been transformed from arcseconds to
Kpcs using the cosmological parameters given in Section~\ref{intro}.

It is worth stressing that just a few dozens of galaxies per clusters,
out of the several hundreds for which WINGS provides surface photometry
parameters, can be included in the final sample, due to the
morphological constraint (early-type) and the cross matching with the
available velocity dispersion data.

\begin{figure}
\epsscale{1.0}
\plotone{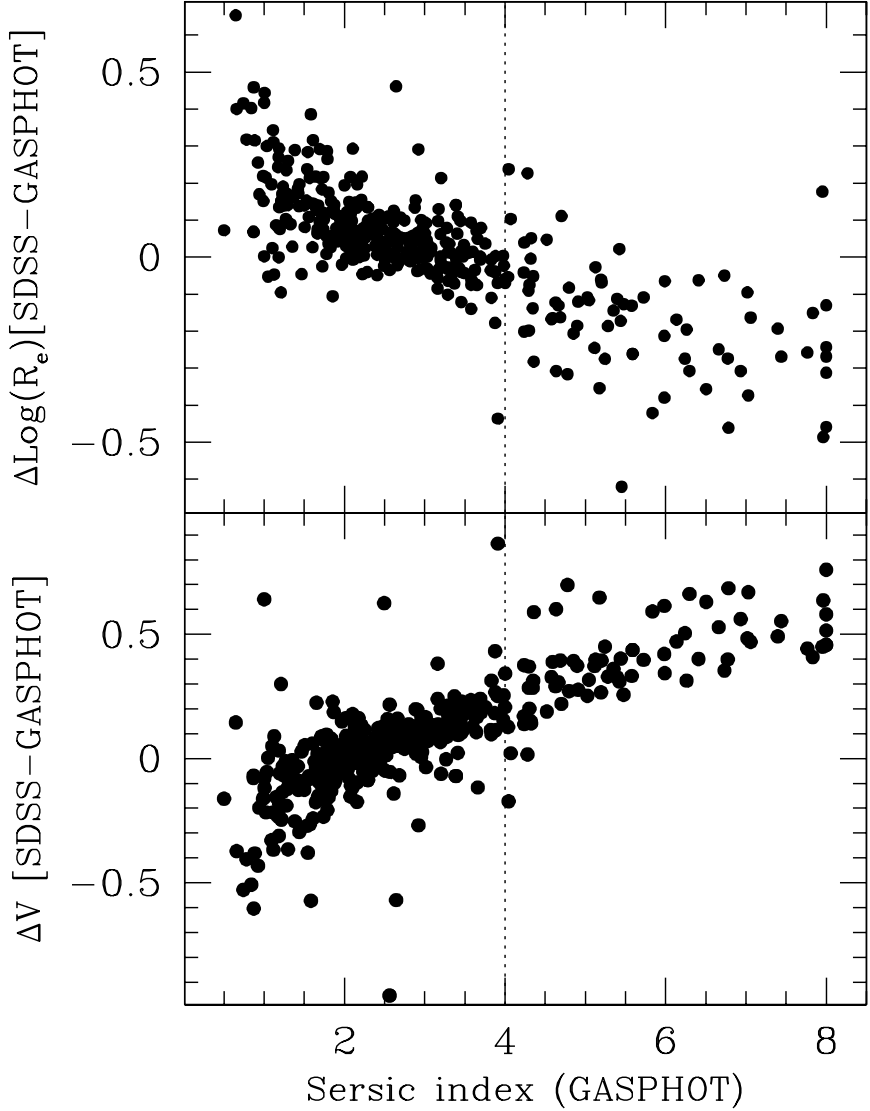}
\caption{
(Upper panel): difference between the effective radii derived by the
SDSS surface photometry using the de~Vaucouleur's \rq\ law and those
derived by GASPHOT using the Sersic's law, as a function of the Sersic
index $n$ (by GASPHOT), for the $407$ ETGs for which both the SDSS and 
the WINGS surveys provide surface photometry parameters.
(Lower panel): as in the upper panel, but for the total V-band
magnitudes. In this case, the SDSS V-band magnitudes are obtained
from the $r'$-band ones using the conversion formula proposed by
\citealp{Fukugita}.
\label{fig1}}
\end{figure}

In Figure~\ref{fig1} we compare total magnitudes and effective radii
derived by GASPHOT (Sersic's law fitting) with the corresponding
quantities derived by the SDSS surface photometry using the
de~Vaucouleur's \rq\ law. The $407$ ETGs in common between the SDSS
and WINGS surveys (including galaxies with $\sigma<95$ or $M_V>-18$)
are shared among $14$ different clusters. The figure clearly
illustrates how the surface photometry parameters are strongly
influenced by the adopted fitting procedure. In particular, in our
case, the strong dependence of both $\Delta$Log(R$_e$) and $\Delta$V
on the Sersic index $n$ is largely expected due to the different
amount of light gathered in the outer luminosity profiles by the
R$^{1/4}$ and Sersic law extrapolations. However, it is worth noting
in Figure~\ref{fig1} that, even for $n$=4 (dotted lines in the figure)
the GASPHOT and SDSS surface photometries give different results, the
last one producing slightly fainter and smaller galaxies. 
To this concern, according to the SDSS-DR6 documentation, both the
effective radii and the total luminosities provided by SDSS for
galaxies in crowded fields (as the clusters are) turn out to be more
and more underestimated at increasing the galaxy luminosity. In the
magnitude range typical of our galaxy sample ($\sim$15$<$V$<\sim$18)
we expect these biases to be of the order of -0.05 and 0.05 for
$\Delta$Log(R$_e$) and $\Delta$V, respectively. While for
$\Delta$Log(R$_e$) the expected bias could be enough in order to
explain the discrepancy in the figure (upper panel), for $\Delta$V
(lower panel) it would be largely insufficient. The residual
discrepancy ($\Delta$V$\sim$0.15) is likely attributable to the
difference between the fitting algorithms used by SDSS (2D - pixel by pixel)
and GASPHOT (major and minor axis growth profiles; see Pignatelli et
al. 2006 for a discussion of the advantages of this fitting 
procedure).

\subsection{The kinematical data}\label{sec12}

The central velocity dispersions $\sigma$ of the ETGs have been taken
from the published data of the NFPS and SDSS--DR6 surveys. It follows
that the completeness is strongly affected by the selection criteria
adopted in these surveys. In particular, the SDSS survey defines ETGs
those objects having both a concentration index $R_{90}/R_{50} > 2.5$
(in the $i^*$ band) and a very good \rq\ de~Vaucouleurs light profile,
while the ETGs of the NFPS survey have been selected on the basis of
their colors, using a narrow strip around the color-magnitude
diagram. Both criteria might lead to exclude from the samples the
brightest cluster galaxies (BCGs), which are actually laking in the
SDSS sample. Moreover, in the SDSS survey the velocity dispersions are
measured only for spectra with signal-to-noise ratio $S/N>10$ (high
average surface brightness) and some clusters are not
fully mapped by the survey strips. We will see that such different
selection criteria produce systematic differences in the
FP coefficients derived for the two samples.

\begin{figure}
\plotone{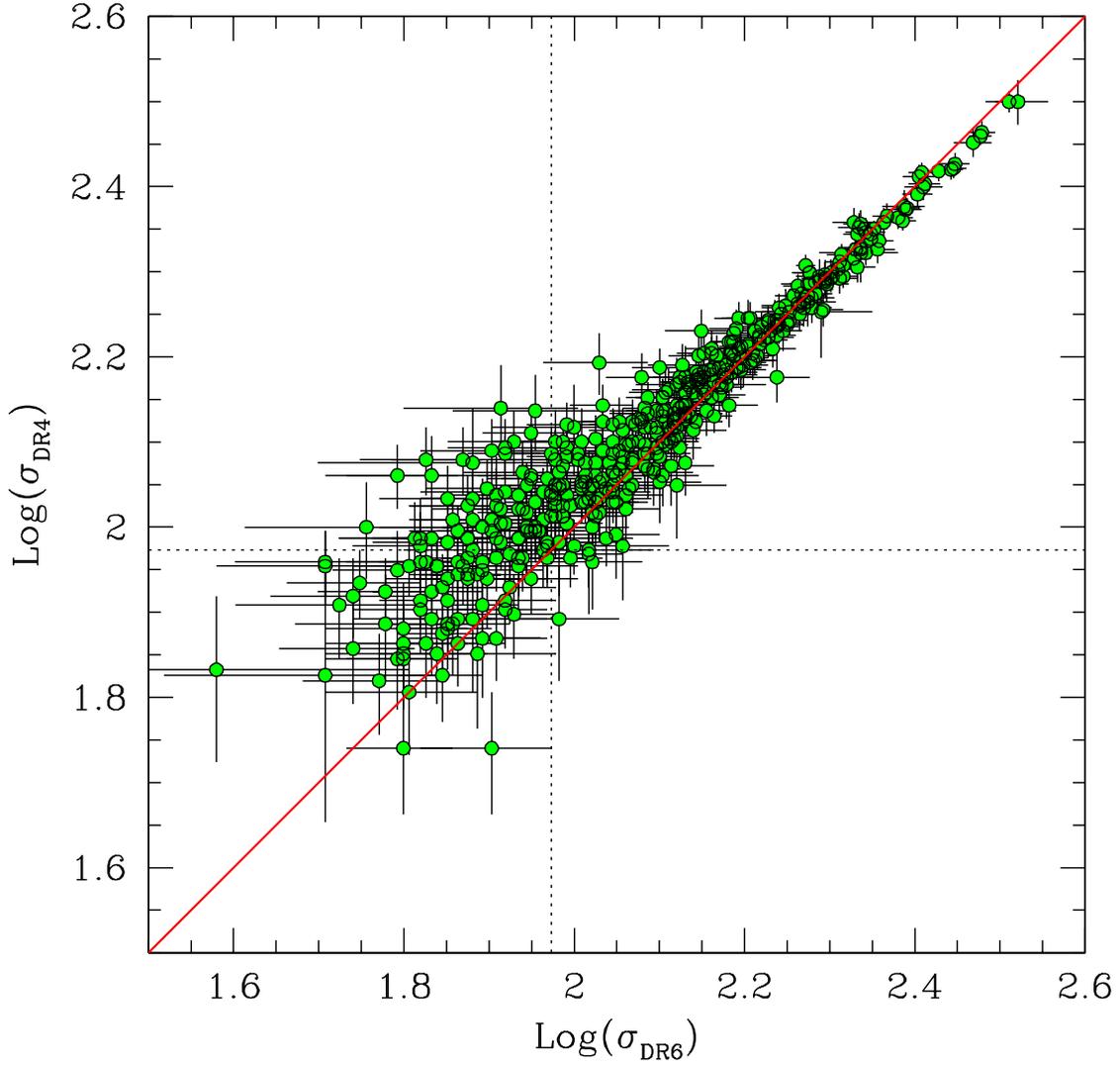}
\caption{ 
Comparison between Log($\sigma_{DR6}$) and Log($\sigma_{DR4}$)
for 523 SDSS galaxies originally selected in the fields of our
WINGS survey. Note the systematic offset between the DR4 and DR6
releases at low velocity dispersions.
\label{fig2}}
\end{figure}

It is worth pointing out that in the originally submitted version of
this paper ({\tt arXiv0804.1892D}) we used SDSS velocity dispersion
data from a previous release of the survey (SDSS--DR4) and that the
differences between the velocity dispersions given in DR4 and DR6 are
not negligible, especially for small values of $\sigma$ (see
Figure~\ref{fig2}). This is the reason why many figures and tables,
as well as some findings we report here (mainly concerning the
difference between the FP coefficients of the SDSS and NFPS samples)
are slightly different from the corresponding ones reported in the
previous version of the paper. Still, we decided to keep that version
unchanged on the babbage (just slightly modifying the title) in order
to show how much a correct determination of the physical quantities
involved in the FP (especially $\sigma$) is critical in drawing any
conclusion from the FP tool.

All the available velocity dispersions have been homogenized to the
uniform aperture $R_e/8$, following the recipe of \citet{Jorg95}. The
estimated uncertainty for both surveys is in the range $7\div10$\%.

\begin{figure}
\plotone{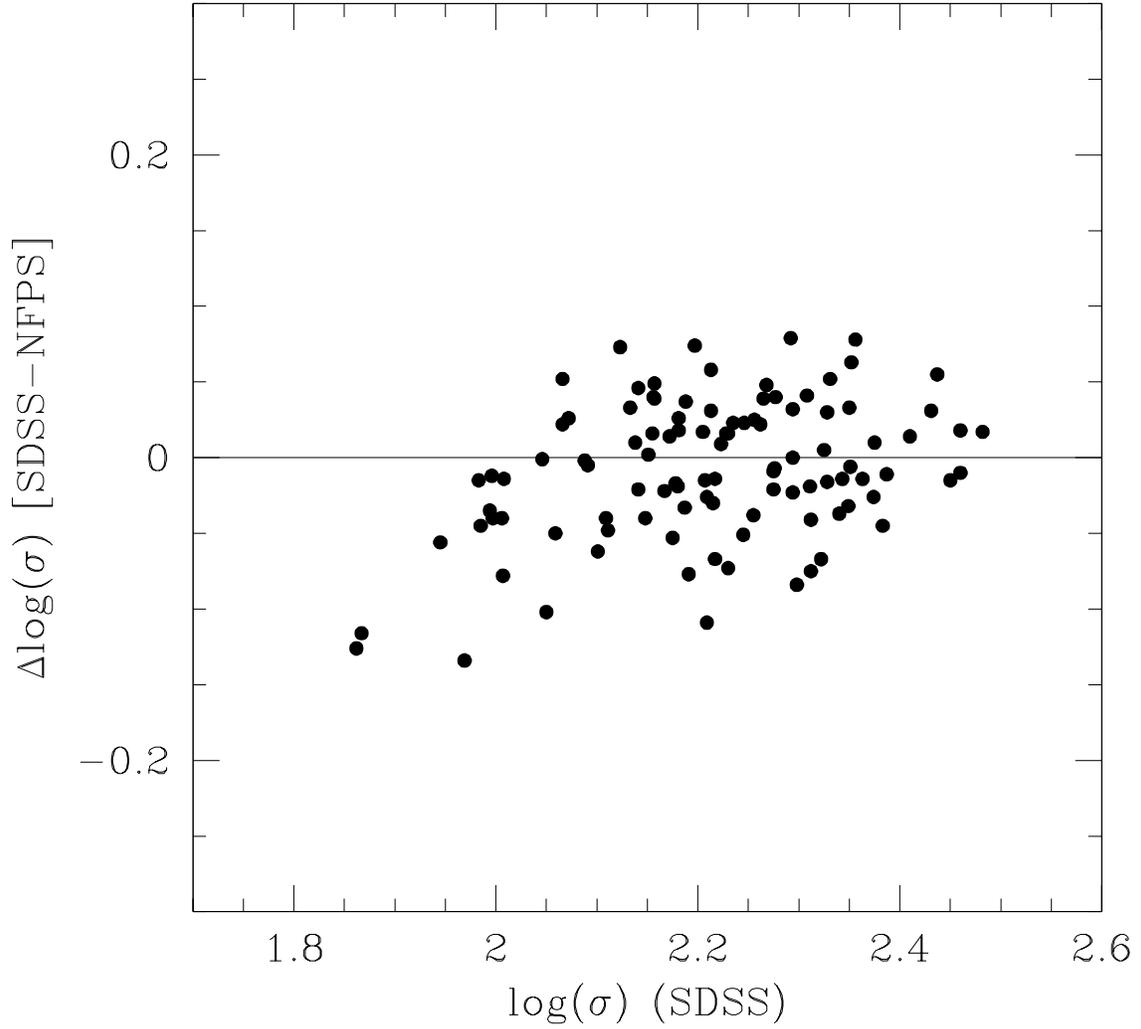}
\caption{ 
The difference Log($\sigma_S$)--Log($\sigma_N$) versus Log($\sigma_S$)
for the 100 ETGs of our sample in common between the NFPS and SDSS
surveys. Note the systematic offset at low velocity dispersions, which
led us to restrict our sample to galaxies with $\sigma>95\ km\ s^{-1}$.
\label{fig3}}
\end{figure}

In Figure~\ref{fig3} we plot the difference
Log($\sigma_N$)--Log($\sigma_S$) versus Log($\sigma_N$) for the $100$
galaxies of our sample in common between the NFPS and SDSS
samples. The $rms$ scatter of the Log($\sigma_N$) vs Log($\sigma_S$)
relation is $\sim 0.05$, equivalent to an uncertainty of $\sim 12$\%
in the common velocity dispersions. Again there is a systematic
deviation between the two datasets at low velocity dispersions
($\sigma<95\ km\ s^{-1}$).

In the following, to avoid any possible bias in the comparison of the
FP of clusters, we have excluded from our analysis the objects with
$\sigma<95\ km\ s^{-1}$.  Moreover, when dealing with the global
(W+N+S) galaxy sample, the average velocity dispersion
$\sigma=(\sigma_N+\sigma_S)/2$ have been assigned to the galaxies in
common between NFPS and SDSS.

\section{FITTING THE FP}\label{sec2}

It is well known that the values of the FP coefficients vary
systematically at varying the adopted fitting algorithm
\citep{Strauss,Blakeslee} and that the choice of the
algorithm actually depends on the particular issue under investigation
(relation among the physical quantities, linear regression for
distance determination, etc..). Here we tried two different algorithms
to get the best fit of the FP: 1) the program MIST, kindly provided by
\citet{LaBarb}, which is a bisector least square fit, coupled with a
bootstrap analysis providing a statistical estimate of the errors of
the FP coefficients; 2) a standard $\chi^2$ fit minimizing the
weighted sum of the orthogonal distances (ORTH hereafter). Both
algorithms account, in different ways, for the measurement errors on
the variables Log($R_e$), \muem, and Log($\sigma$). MIST considers an
average covariance matrix that includes the variances of the errors in
all parameters and their mutual correlations (such as $rms_{Log(R_e)}$
vs. $rms_{\muem}$).  On the other hand, ORTH takes into account the
errors of individual measures in a standard $\chi^2$ analysis.  In
Table~\ref{tbl2} we report the FP coefficients derived from the two
fitting algorithms for the global galaxy sample (first two lines) and
for the NFPS and SDSS samples separately (lines 3-4 and 5-6,
respectively). In the same table (lines 8-9) we report for comparison
the FP coefficients obtained with both MIST and ORTH fitting
algorithms for a sample of 80 ETGs in the Coma cluster (photometric
and kinematical data from \citealp{Jorg95}[JORG]). The column labeled
with $N_g$ in the table reports the number of galaxies used in each
fit.

Besides the best fitting algorithm, the FP coefficients might also be
systematically influenced by the technique adopted to measure the
effective radius and surface brightness of galaxies (1D/2D light
profile fitting with de~Vaucouleurs/Sersic laws). Lines 5 and 7 of
Table~\ref{tbl2} report the MIST FP coefficients obtained for the
galaxy sample in common between WINGS and SDSS, using alternatively
the two surface photometry data sets (see in Figure~\ref{fig1} the
comparison among them and in Section~\ref{sec11} the description of
the WINGS and SDSS surface photometry techniques). It is evident that,
at least in our case, the influence of the adopted surface photometry 
technique on the FP coefficients turns out to be negligible.

\placetable{tbl2}

Table~\ref{tbl2} shows that different fitting algorithms (and,
possibly, surface photometry techniques) lead to somewhat systematic
differences in the FP coefficients. In particular, the values of $a$
obtained using the MIST fit are in general slightly smaller than those
coming from the orthogonal fit. This means that, in order to perform a
correct comparison of the FP results, it is advisable to adopt
homogeneous FP fitting and (perhaps) surface photometry techniques.

However, in the present analysis, we do not focus on the 'true' values
of the FP coefficients. Instead, we will concentrate on their possible
variation as a function of both galaxy and cluster properties. In
other words, rather than in obtaining the best possible fit for a
given application of the FP, we are interested in investigating the FP
systematics, once both the fitting algorithm and the surface
photometry technique have been chosen. Hereafter we adopt the MIST
bisector fitting algorithm and the WINGS-GASPHOT surface photometry.
The last choice will allow us to account for the structural
non-homology of galaxies (Sersic index, see Section~\ref{sec5}), while
the former one will provide FP coefficients useful for distance
determination of farther clusters. However, using the ORTH fitting
algorithm, we will also provide in Section~\ref{sec43} a recipe for
the V-band FP, useful to define the physical relation among the
quantities involved in it.

Comparing each one another the FP coefficients given in
Table~\ref{tbl2}, we easily realize that, besides the obvious
dependence on the fitting algorithms, a further dependence exists on
the galaxy sample, even adopting the same fitting algorithm (MIST) and
surface photometry technique (WINGS-GASPHOT). In particular, the $a$
coefficient, which is related to the so called 'tilt' of the FP, is
noticeably different for the three data samples, even if the $rms$
scatter in Log($R_e$) is always $\sim 0.05$ (which implies an
uncertainty of $\sim12\%$), a value just a bit larger than that
reported in \citet{Jorg95} ($\sim11\%$).

\begin{figure}
\vspace{-2cm}
\hspace{-0.5cm}
\includegraphics[angle=0,scale=0.85]{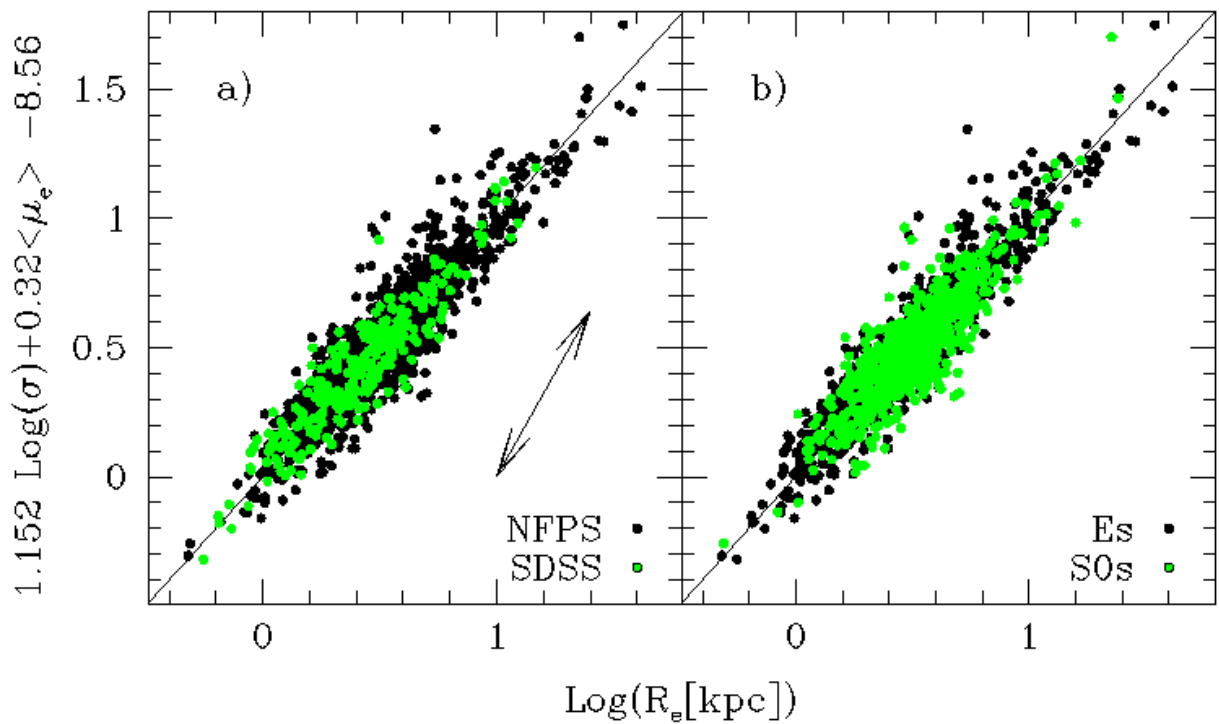}
\caption{ 
(panel a): The FP of the W+N (black dots) and W+S (gray dots; green in
the electronic version) data samples; (panel b) The W+N+S FP for E
(black dots) and S0 (gray dots; green in the electronic version)
galaxies. In both panels we used for reference the FP coefficients
derived from the best fit of the global dataset (W+N+S). 
The two-sided arrow in panel (a) roughly defines,
through the Faber-Jackson (L$-\sigma$) relation, the direction of
constant luminosity (or $\sigma$; see Section~\ref{sec32}).
\label{fig4}}
\end{figure}

In panel (a) of Figure~\ref{fig4} we show the MIST bisector fit of the
FP obtained for the whole W+N+S dataset (see line 1 of
Table~\ref{tbl2}) using two different colors for the W+N and W+S data
samples (respectively black and grey; green in the electronic
form). Note the cut shown by SDSS data at large values of Log($R_e$),
which is obviously due to the bright end cut of the survey. 

\placetable{tbl3}

\begin{figure}
\plotone{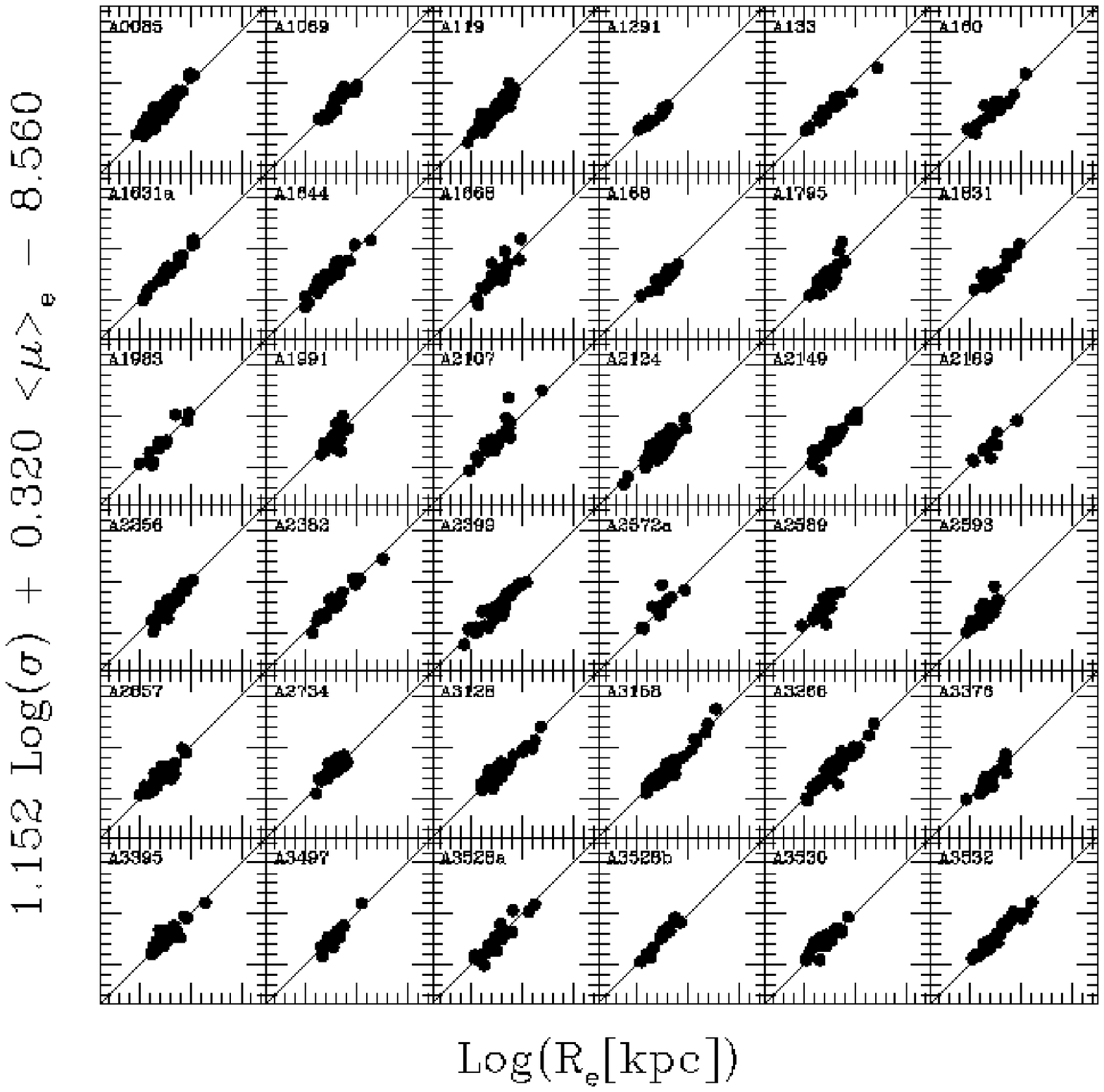}
\caption{The FPs of individual clusters in our sample are plotted 
using the MIST best fit solution found for the global W+N+S galaxy sample.
\label{fig5}}
\end{figure}

In Figure~\ref{fig5} we plot the FP of the individual clusters, again
using for reference the coefficients derived from the fit of the whole
W+N+S data sample. Columns 3-8 of Table~\ref{tbl3} report the best fit
coefficients (and the associated uncertainties) of each cluster,
obtained with the MIST algorithm. Even from a quick look of both
Figure~\ref{fig5} and Table~\ref{tbl3}, it is clear that the global fit
does not seem to be a valid solution for all clusters.

\addtocounter{figure}{-1}
\begin{figure}
\plotone{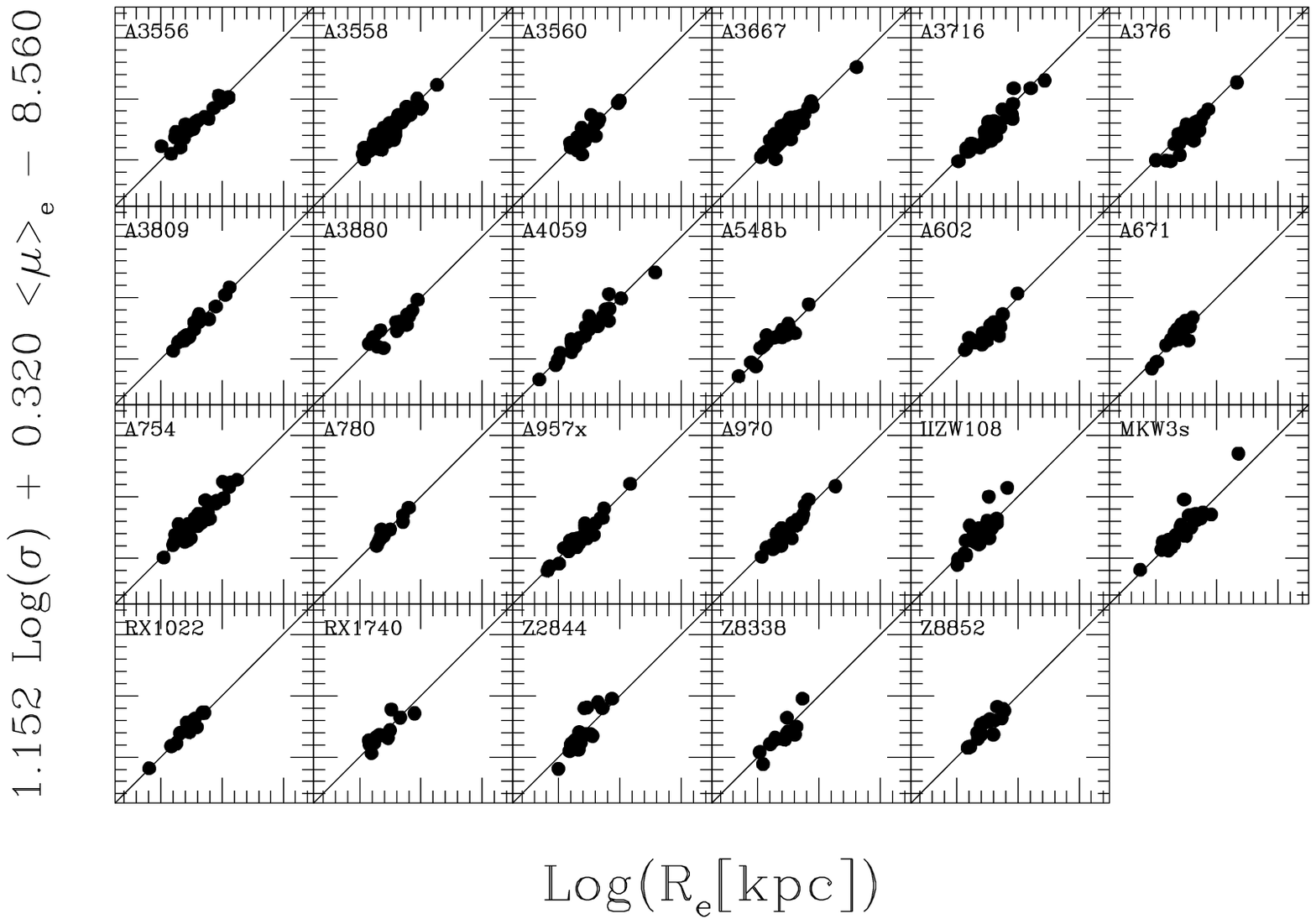}
\caption{The FPs of individual clusters (continued).}
\end{figure}

The average values (with their uncertainties), the standard deviations
and the median values of the (MIST) FP coefficients of the clusters in
the global sample and in the samples W+N and W+S are reported in
Table.~\ref{tbl4}. From this table we note that: (i) even if the
scatter is large, the average values of the FP coefficients appear
systematically different (well beyond the expected uncertainties) in
the W+N and W+S cluster samples, confirming the dichotomy already
noted in Table~\ref{tbl2}; (ii) when just clusters with
$N_g>N_{med}(=23)$ are considered, the standard deviations of the
distributions of the FP coefficients decrease only slightly, suggesting
that the large scatter cannot be ascribed to the statistical
uncertainties related to the (sometimes) small number of ETGs in our
clusters.

\placetable{tbl4}

\section{ORIGIN OF THE SCATTER OF THE FP COEFFICIENTS}\label{sec3}

We test two different hypotheses to explain the differences between
the W+N and W+S samples and, in general, the large observed scatters
of the FP coefficients: (i) they are simply due to the statistical
uncertainties of the fits; (ii) they are artificially produced by the
different criteria used to select ETGs in the NFPS and SDSS surveys.

\subsection{Consistency with statistical uncertainties}\label{sec31}

\begin{figure}
\plotone{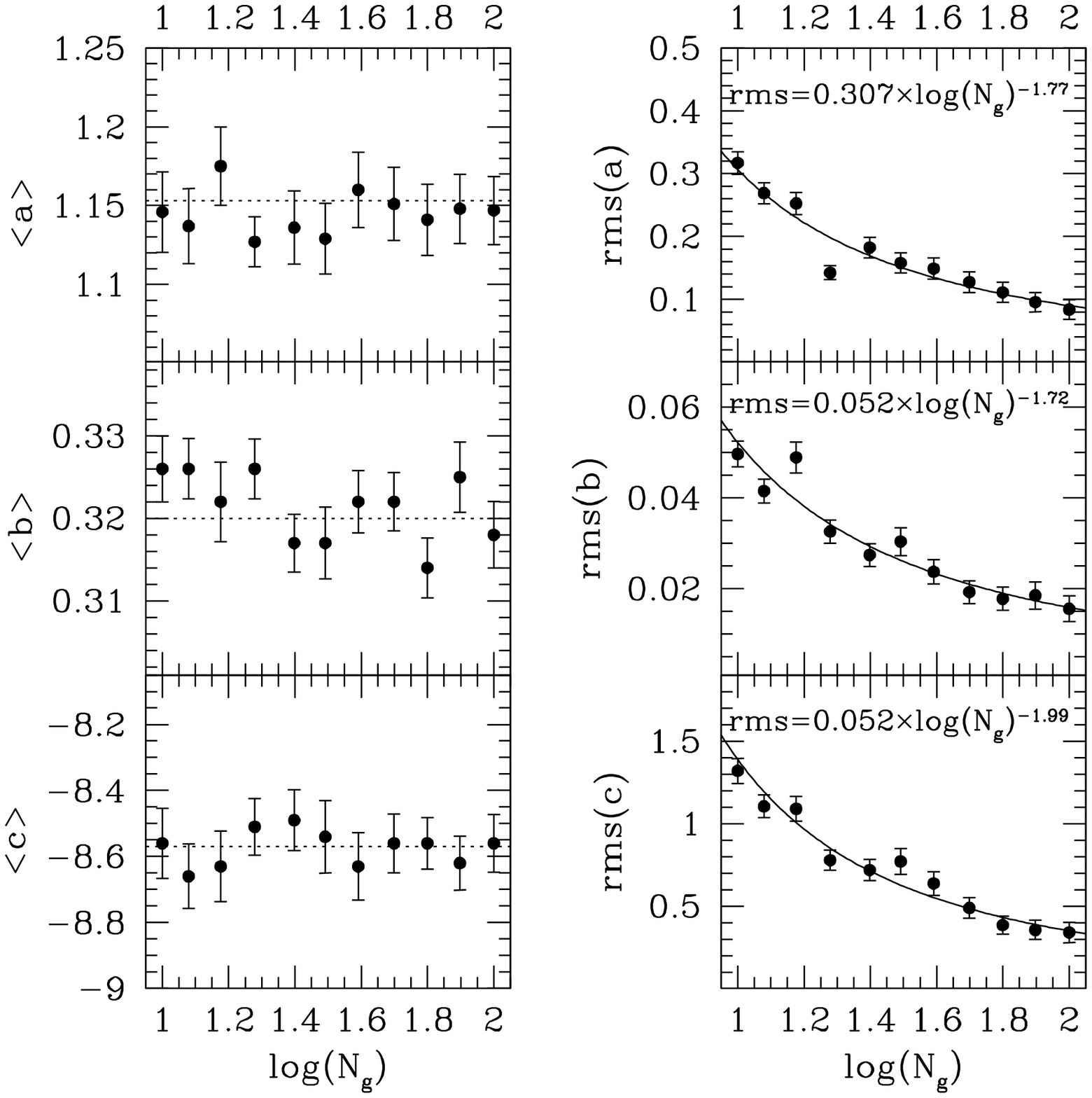}
\caption{
Average values (left panels) and standard deviations (right panels) of
the FP coefficients as a function of the number of galaxies ($N_g$)
for mock clusters randomly extracted from the whole galaxy sample (see
text for details).  The dashed lines in the left panels correspond to
the FP coefficients obtained fitting altogether with MIST the 1550
galaxies in our sample. The full lines in the right panels correspond
to the simple exponential functions we used to compute the standard
deviations as a function of Log($N_g$) (see the equations in each
panel).
\label{fig6}}
\end{figure}

First we test the 'null hypothesis' that the observed scatter is
merely consistent with the statistical uncertainties of the fits. To
this aim, using all the galaxies in our sample, we produced two
different sets of simulated clusters. In the first set we generate
mock clusters with number of galaxies ($N_g$) progressively increasing
from Log($N_g$) = 1 to 2 (step 0.1) and fit each mock cluster with
MIST. Figure~\ref{fig6} shows the average values of the FP coefficients
and the corresponding standard deviations as a function of
Log($N_g$). Note that, since for each value of $N_g$ the whole sample 
of 1550 galaxies is used to randomly extract as many mock
clusters as possible avoiding galaxy repetitions, the number of mock
clusters increases at decreasing $N_g$, thus resulting in almost
constant error bars of the average FP coefficients and of their
variances.

In the second set of simulations we produced 100 toy surveys, each
containing 59 clusters obtained by sorting randomly the whole galaxy
sample and taking sequentially the same number of galaxies per cluster
as the real survey (thus avoiding galaxy repetitions)\footnote{Note
that, in this way, we implicitly assume that the probability distributions 
of photometric/kinematic properties of
galaxies are the same in all clusters and correspond to those of the
global galaxy sample.}. Then, using the MIST algorithm, we evaluate
the FP coefficients of each mock cluster and, for each mock survey, we
compute the average and median values of the coefficients, together
with their standard deviations. Finally, we compare the distributions
of the average coefficients and their variances in the mock surveys
with the corresponding values of the real survey.

\begin{figure}
\vspace{-2cm}
\hspace{-0.5cm}
\includegraphics[scale=0.8]{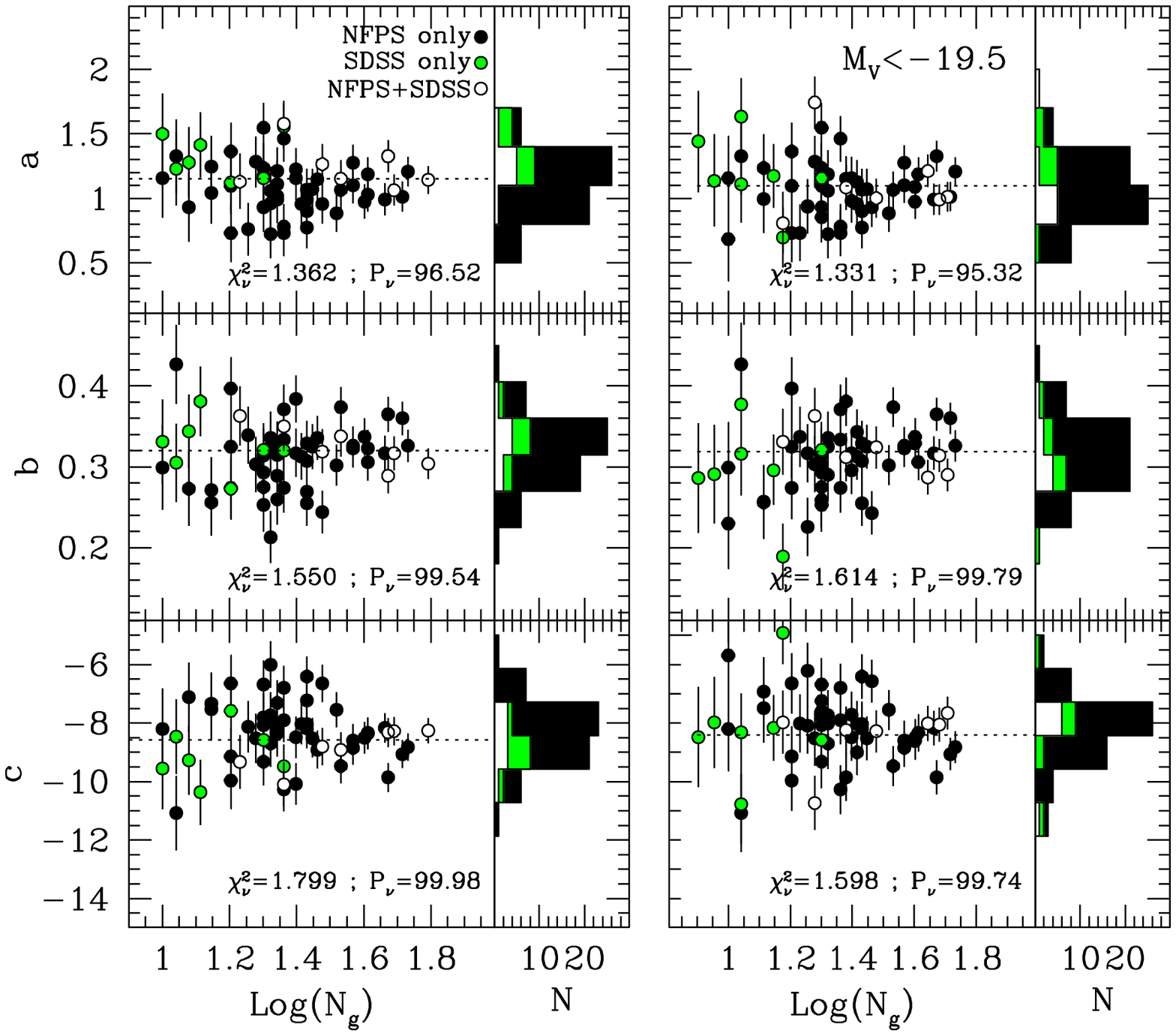}
\vspace{-1.5cm}
\caption{
(Left panels): FP coefficients of our clusters versus the number of
galaxies in each cluster. The black and gray (green in the electronic
version) dots refer to clusters with only NFPS and SDSS galaxies, respectively.
The open dots represent the seven clusters in common between the NFPS
and SDSS surveys. The dashed lines correspond to the FP coefficients
obtained fitting altogether with MIST the 1550 galaxies in our
sample. The histograms on the right represent the
distributions of FP coefficients in our cluster samples. Black, gray
(green in the electronic version) and open histograms have the same
meanings as in the left plots and are cumulated inside each bin;
(Right panels): as in the left panels, but using only galaxies with
$M_V<-19.5$.  Note that in the left panels (global galaxy sample) the
NFPS and SDSS clusters have quite different distributions of the 
coefficients, while in the right panels (just galaxies with $M_V<-19.5$) 
the distributions of the two samples are consistent among each
other.
\label{fig7}}
\end{figure}

\begin{figure}
\plotone{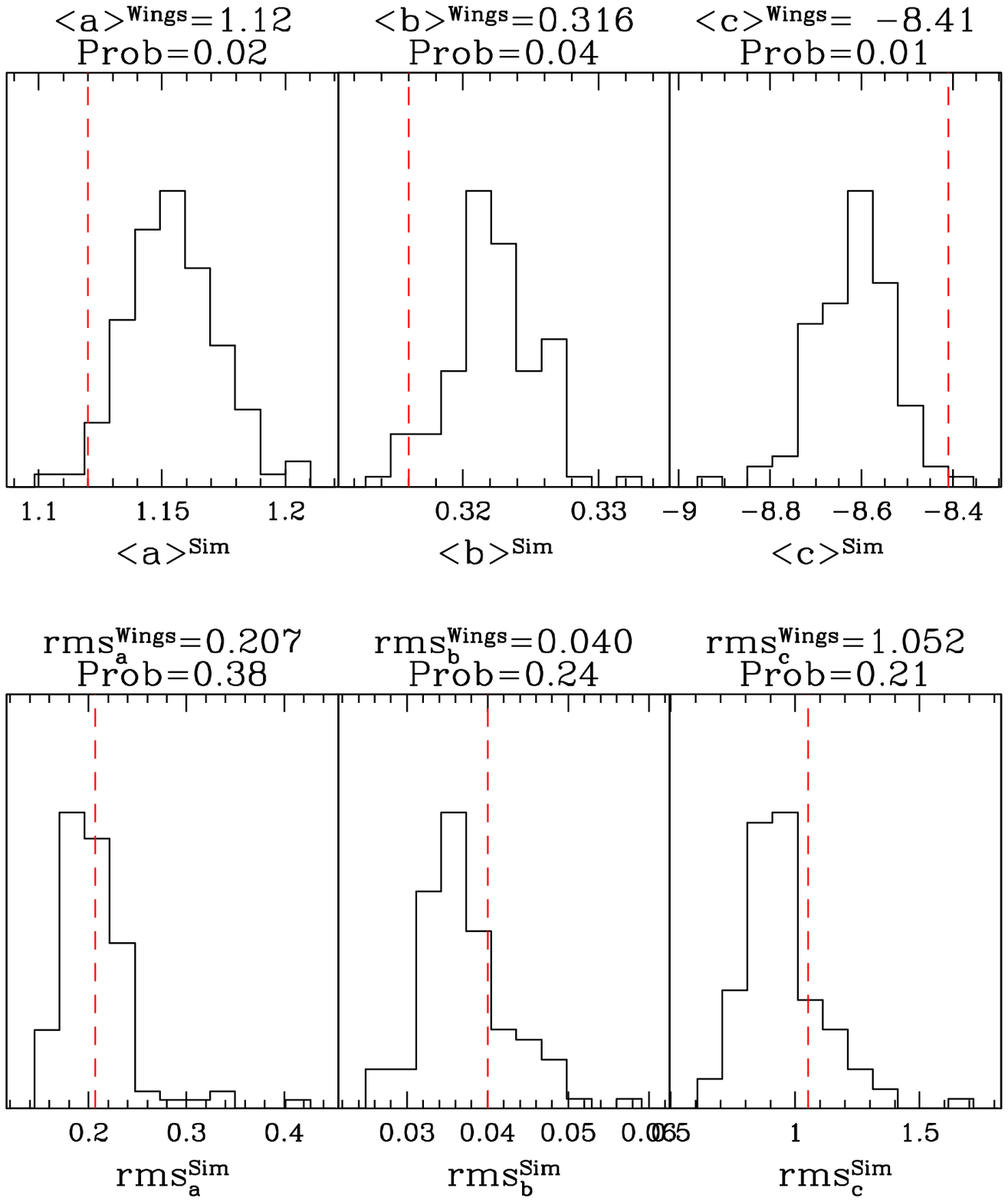}
\caption{
Histograms of the average values of the FP coefficients (upper panels)
and standard deviations (lower panels) for the 100 toy surveys (see
text for details). The dashed lines in the histograms mark the
corresponding values obtained from the real survey (see
Table.~\ref{tbl4}). These values are also reported in the panels,
together with the probabilities that they are randomly extracted from
the underlying histograms.  Note in the upper panels that the average
values of the FP coefficients for the clusters of the real
survey are just marginally consistent with the corresponding distributions
obtained from the mock surveys.
\label{fig8}}
\end{figure}

Figures~\ref{fig7} and~\ref{fig8} illustrate the conclusions of the
two sets of simulations. The first set has been used to
compute (with the equations given in the right panels of
Figure~\ref{fig6}) the error bars in Figure~\ref{fig7}. The left
panels of this figure report the FP coefficients of our 'real'
clusters versus the number of galaxies in each cluster, while the
histograms on the right side of each panel show the corresponding
distributions (see figure caption for more details). The error bars
are used to compute the reduced Chi-square values (reported in the
figure; in our case $\nu$=58) of the differences between the
coefficients of the individual clusters and the corresponding
coefficient of the global galaxy sample (dashed lines in the
figure). Apart from the $a$ coefficient ($P_{\nu}\sim$0.965), they
correspond to very high values of the rejection probability
($P_{\nu}>$0.995) that the coefficients of the individual clusters are
randomly extracted from the same parent population.

Figure~\ref{fig8} shows that the average values of the FP
coefficients for the clusters of the real survey are just marginally 
consistent with the corresponding distributions obtained with the 
simulations of mock surveys (upper panels), while the distributions of 
variances are more or less in agreement with the real ones (lower
panels).

The two sets of simulations indicate that the observed scatter is not
accounted for by the statistical uncertainties of the fits and that
the real clusters cannot be merely assembled by random extraction of
galaxies from the global population. The left panels of
Figure~\ref{fig7} also clearly illustrate the systematic differences
between the FP coefficients of the NFPS and SDSS samples already
quoted in Tables~\ref{tbl2} and \ref{tbl4}. To this concern, the
two--sample Kolmogorov--Smirnov test, applied to the black and
open+gray(green) samples in that Figure, provides rejection probabilities of
0.998, 0.530 and 0.986 for the left panels coefficients $a$, $b$ and
$c$, respectively.

\subsection{Dependence on galaxy sampling}\label{sec32}

As an early test, we wanted to investigate the hypothesis that the
observed scatter of differences in the FP coefficients are the result
of blending E and S0 galaxies, with the knowledge that the E/S0 ratio
varies (for example) as a function of local density. We verified that
the FP computed separately for the elliptical and S0 galaxies are
practically indistinguishable (all FP coefficients differ by $<3\%$;
see also the right panels of Figure~\ref{fig4}). This allows us to rule
out the hypothesis that the observed scatter is induced by different
E/S0 fractions in the different samples.

On the other hand, we have seen in Tables~\ref{tbl2} and~\ref{tbl4}
that the FP coefficients of the clusters in the W+N and W+S data
samples are systematically different from each other (see also the
left panels of Figure~\ref{fig7} and the last sentence of the previous
sub--section). It is therefore natural asking which is the origin of
such systematic difference.

Since for all galaxies in the sample the surface photometry data come
from WINGS+GASPHOT, we could be tempted to conclude that the
differences we found are due to some systematic offset between
velocity dispersion measurements from the NFPS and SDSS
surveys. However, this possibility is definitely ruled out by
Figure~\ref{fig3} (Sec.~\ref{sec12}), which shows that the agreement
between the two velocity dispersion surveys is fairly good, at least 
for $\sigma >$95km~s$^{-1}$. Indeed, we
have also verified that the FP coefficients of the galaxy sample in
common between NFPS and SDSS, obtained using alternatively the two
velocity dispersion data sets do not differ significantly.

\begin{figure}
\plotone{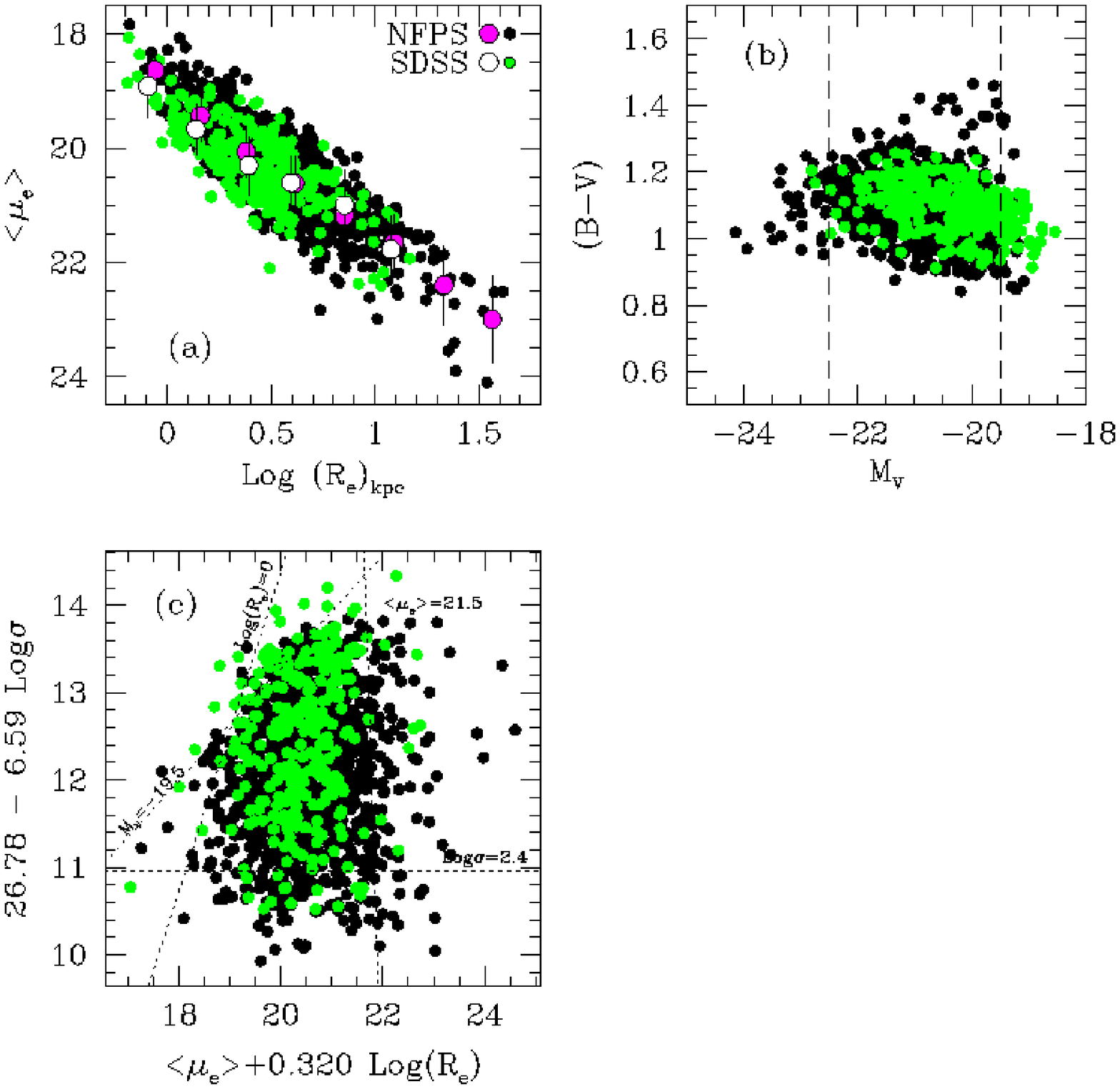}
\caption{
(Panel a): the \muere\ relation for the W+N (black dots) and W+S (gray
dots; green in the electronic version) samples. The big dark-gray
(magenta in the electronic version) and white dots represent the
average surface brightnesses of the two galaxy samples in different
bins of Log($R_e$); (Panel b): the Color-Magnitude diagrams for the
galaxies of the W+N and W+S samples. Symbols are as in panel (a);
(Panel c): face-on view of the FP obtained for the global galaxy
sample. The dotted straight lines mark the loci corresponding to some
constant values of luminosity, surface brightness, effective radius
and velocity dispersion. Symbols as in the previous panels.
\label{fig9}}
\end{figure}

Thus, we are left with the last possibility: that the systematic FP
differences between the NFPS and SDSS clusters are due to the
different distributions of photometric/kinematic properties of
galaxies in the two samples. The danger of selection biases in this
game has already been emphasized by \citet{Lynden-Bell},
\citet{Scodeggio}, and \citet{sloan}, who showed that robust fits of
the FP can be obtained only for galaxy samples complete in luminosity,
volume, cluster area coverage and stellar kinematics. The panels (a)
and (b) of Figure~\ref{fig9} respectively show the projection of the
FP on the surface photometry plane (\muere\ ; Kormendy relation) and the
Color-Magnitude diagrams [$M_V - (B-V)$] for the NFPS and SDSS
surveys. Both figures show that the two galaxy samples have quite
different distributions of the photometric quantities involved in the
FP parameters. This is even more evident in the panel (c) of the same
figure, where the face-on view of the FP of the global sample is
shown, together with the loci corresponding to some constant values of
the quantities involved in the FP (dotted lines). Note that, both in
the Color-Magnitude and in the Kormendy diagrams, the W+S galaxy
sample turns out to be (on average) fainter than the W+N sample,
especially in the small size region. This is likely a direct
consequence of the rules the two surveys adopt to select early-type
galaxies (see Section~\ref{sec12}).

\begin{figure}
\vspace{-4cm}
\hspace{-1cm}
\includegraphics[angle=-90,scale=0.65]{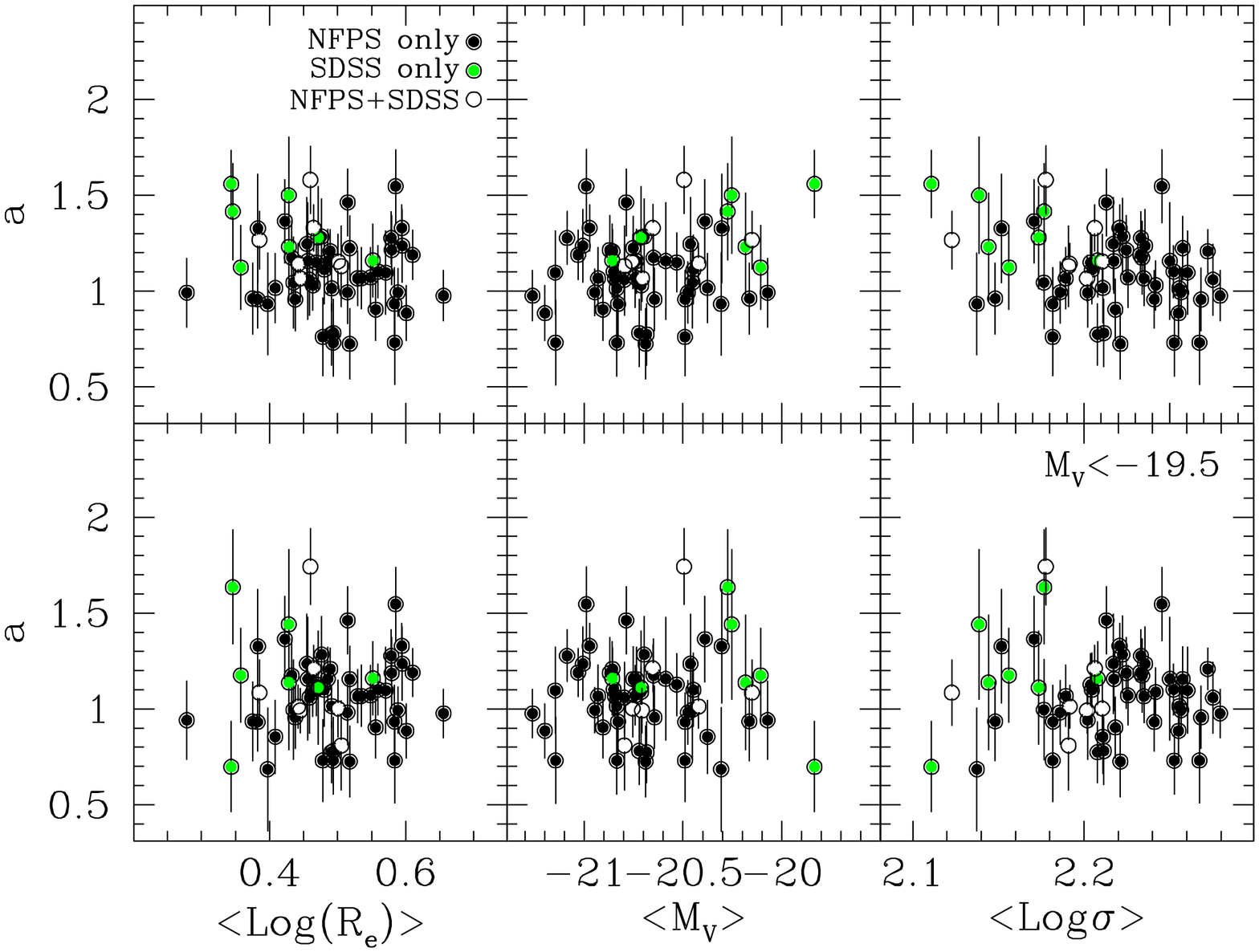}
\vspace{0.5cm}
\caption{(Upper panels) the FP coefficient $a$ vs. the average values of
effective radius ($<Log(\re)>$), luminosity ($<M_V>$) and central
velocity dispersion ($<Log(\sigma)>$) of the galaxies in each
cluster. Symbols are as in Figure~\ref{fig7}. (Lower panels) the
same plots, but using only galaxies with $M_V<-19.5$.
\label{fig10}}
\end{figure}

The fact that such differences in the galaxy sampling produce the
observed differences in the FP coefficients is shown in
Figure~\ref{fig10}. In the upper panels of the figure the $a$ coefficient
of the FP seems to be anti-correlated with the average values of luminosity,
radius and velocity dispersions of galaxies in the clusters. The same,
but (obviously) with positive CCs, happens for the coefficient $c$ (not reported
in the figure). We see from the lower panels in the figure that, if we
cut the data samples at higher luminosity, $M_V=-19.5$, these
correlations disappear, since in this case the two data samples
are more homogenous.

This is also confirmed by the right panels of Figure~\ref{fig7}, where
the plots in the left panels are repeated using only galaxies with
absolute magnitude $M_V<-19.5$. Indeed, the two--sample
Kolmogorov--Smirnov test, applied to the black and open+gray(green)
samples in that figure, provides rejection probabilities of 0.475,
0.308 and 0.318 for the right panels coefficients $a$, $b$ and $c$,
respectively (compare these values with those given in the last
sentence of Section~\ref{sec31})

\begin{figure}
\includegraphics[angle=-90,scale=0.65]{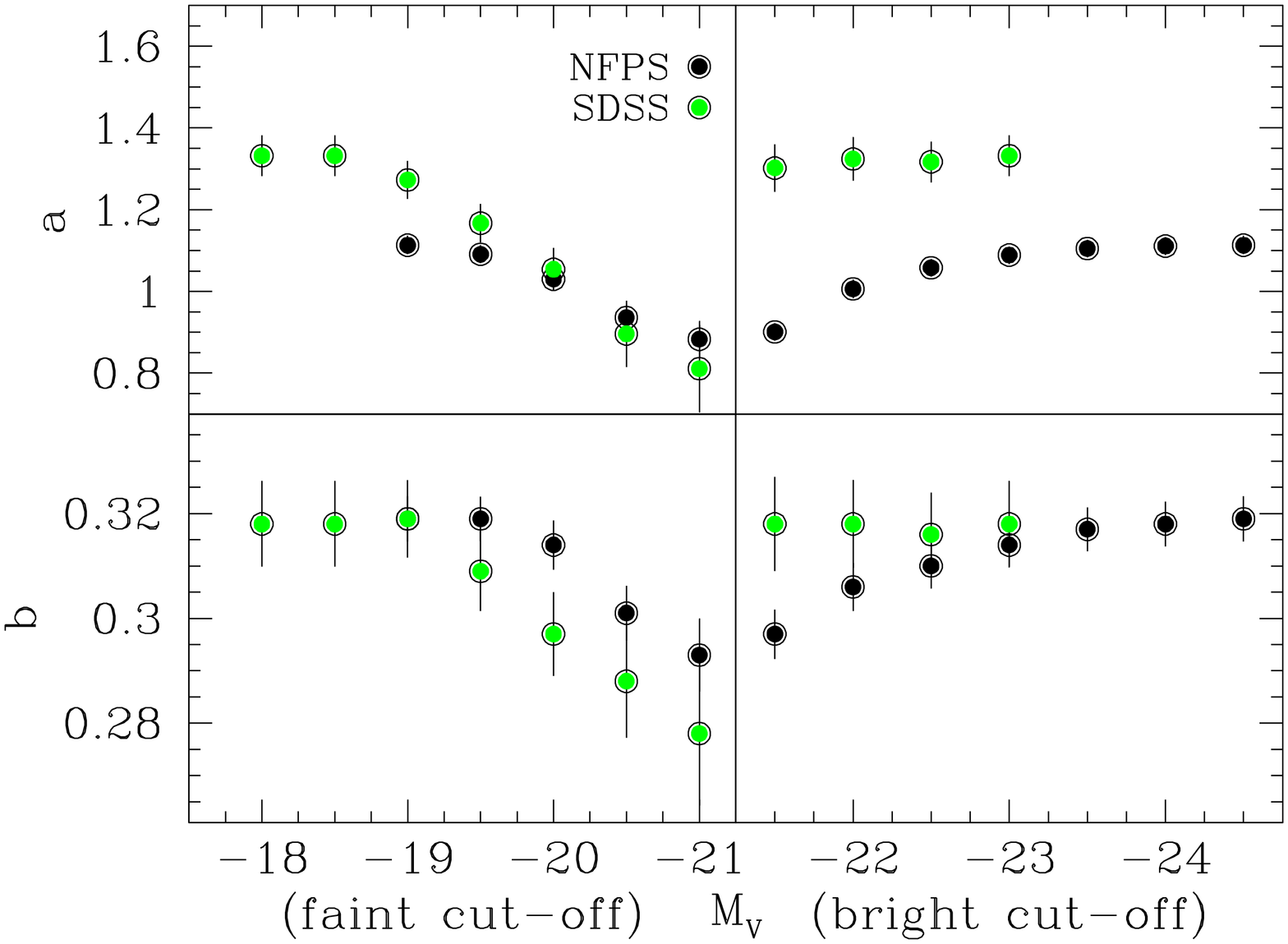}
\vspace{0.5cm}
\caption{The FP coefficients $a$ and $b$ obtained with MIST for different values of the
faint and bright luminosity cut-off applied to the NFPS galaxy sample (black dots) and to
the SDSS sample (gray dots; green in the electronic version).
\label{fig11}}
\end{figure}

A final, quantitative estimate of the dependence of the FP
coefficients on the luminosity distribution of the galaxy
sample is provided by Figure~\ref{fig11}, where we report the FP
coefficients obtained for different values of the faint and
bright luminosity cut-off applied to the NFPS and SDSS
samples.

\begin{figure}
\includegraphics[angle=-90,scale=0.65]{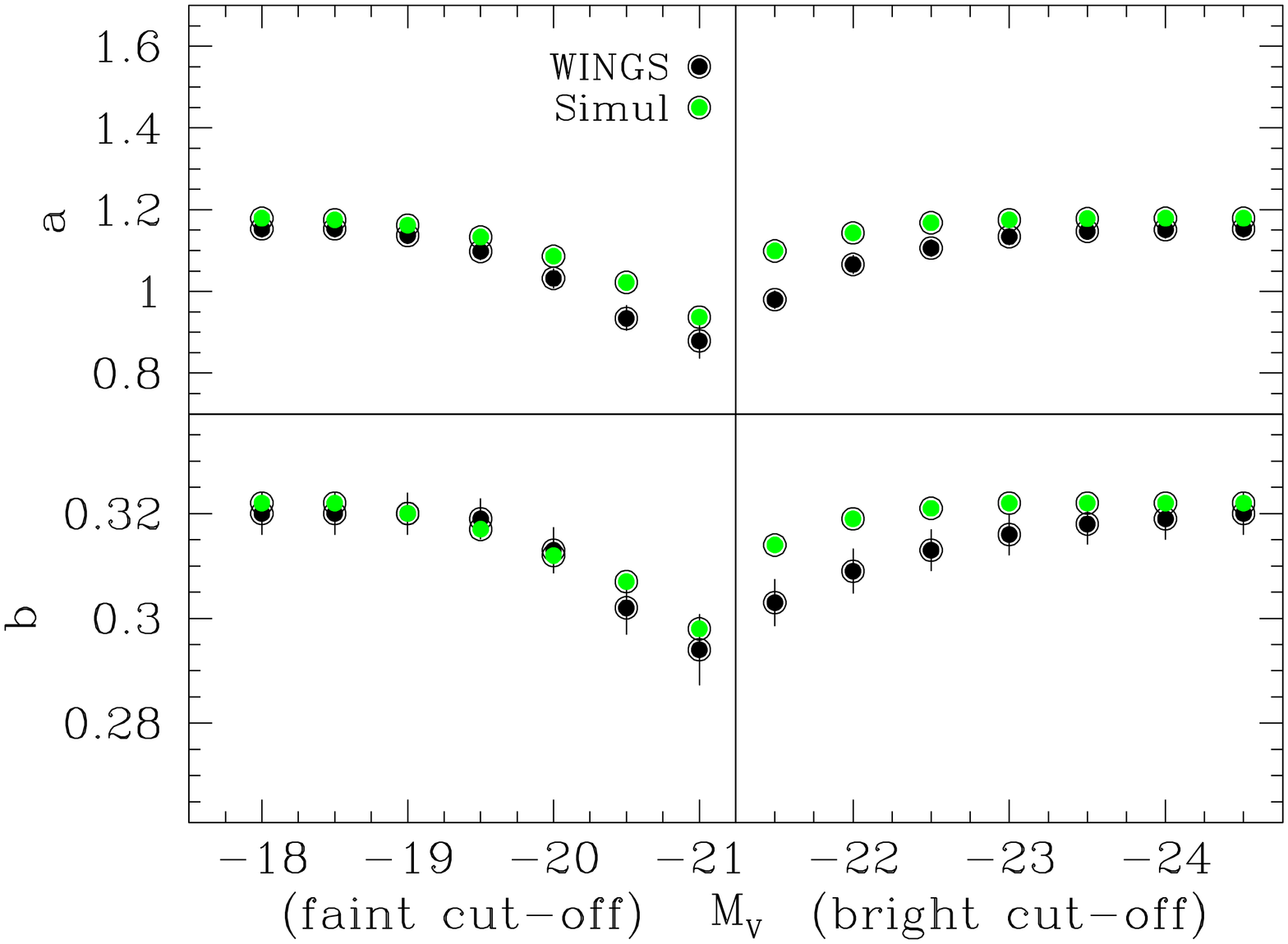}
\vspace{0.5cm}
\caption{similar to Figure~\ref{fig11},
but comparing the W+N+S sample (black dots) with a mock sample of
10,000 toy galaxies (grey dots; green in the electronic version of the
paper; see text for more datails)
\label{fig12}}
\end{figure}

The left panels of Figure~\ref{fig11} show that, for both the NFPS and
the SDSS sample, the coefficients $a$ and $b$ decrease at increasing
the faint luminosity cut. This effect can be at least partially
explained by the very geometry of the FP. In fact, in the edge-on
representation of the FP, any luminosity cut in the galaxy sample
translates, through the Faber-Jackson (L$-\sigma$) relation, in a sort
of 'zone of avoidance' delimited by a line of constant luminosity (or
$\sigma$), whose direction is roughly indicated by the two-sided arrow
in Figure~\ref{fig4} (panel a). This {\it Malmquist-like} bias reduces
the FP slopes along the directions of $\sigma$ and \muem\ for both
faint- and bright-end luminosity cuts. Figure~\ref{fig12} illustrates
this {\it 'geometrical'} effect. It is similar to Figure~\ref{fig11},
but compares the W+N+S sample (black dots) with a mock sample of
10,000 toy galaxies (grey dots; green in the electronic version of the
paper) randomly generated around the same (W+N+S) FP, according to the
'true' distributions (and mutual correlations) of \muem\, \re\ and
$\sigma$. The right panels of Figure~\ref{fig11} show that this bias
actually works for the bright-end luminosity cut just in the
case of the NFPS sample. Instead, the FP coefficients of the SDSS
sample display a rather peculiar behaviour. For the faint-end cuts
they show trends similar to those of the NFPS sample, but more
pronounced. Instead, they do not seem to depend at all on the bright-end
cuts (right panels), remaining significantly higher than in the case
of the NFPS samples over the range of cut-off luminosities. This
behaviour suggests that, besides the luminosity cut-off, other causes
may contribute to tell apart the two samples.

\begin{figure}
\plotone{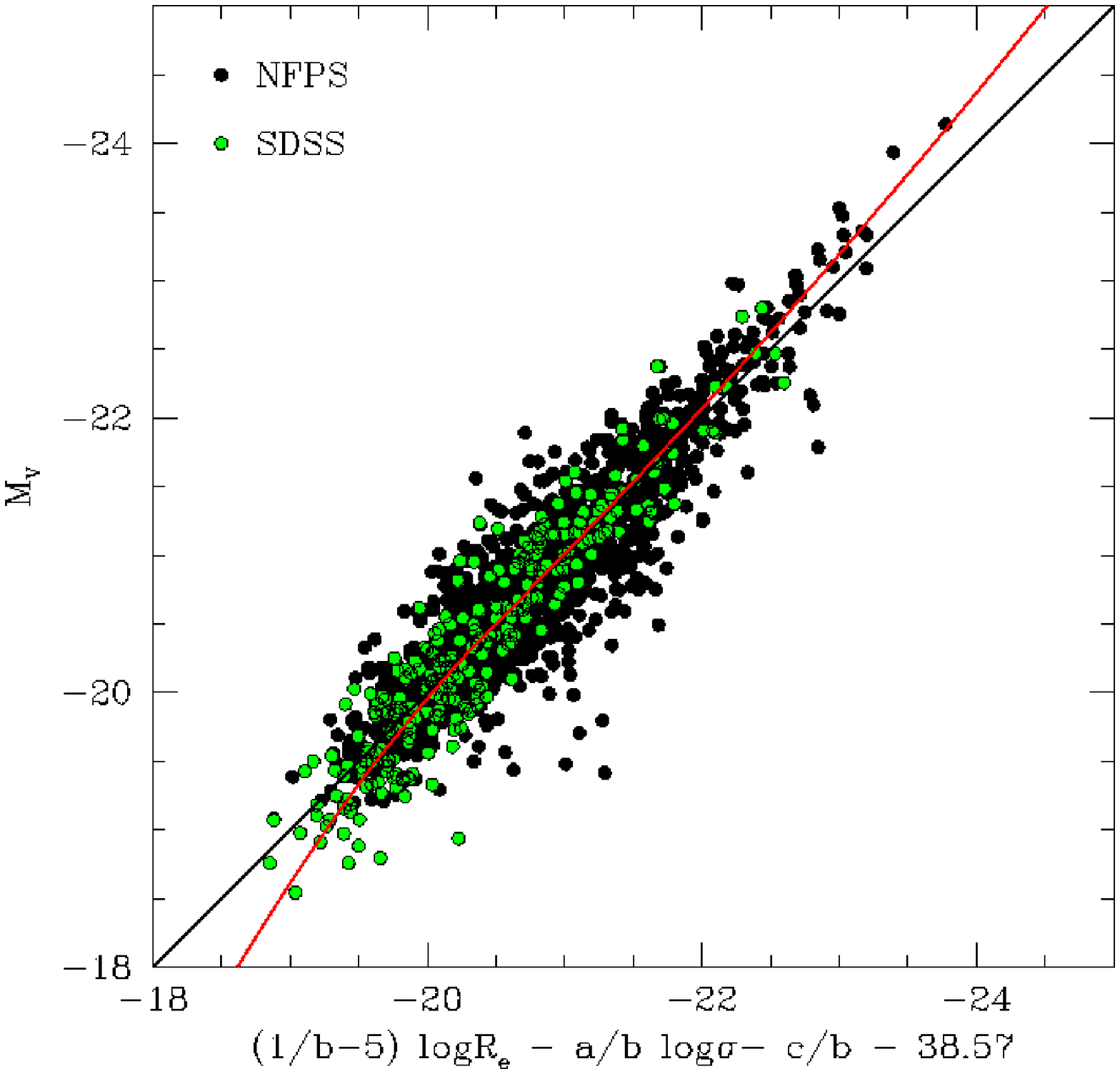}
\caption{
Edge-on FP as it appears along the direction of luminosity. Symbols
are as in panel (a) of Figure~\ref{fig4}. The black curve (red in the
electronic version) just represents a naive fitting we made in order
to enhance the warp-like feature of this particular projection of the FP.
\label{fig13}}
\end{figure}

The Figure~\ref{fig13} helps to clarify this point. It shows the
edge-on FP as it appears along the direction of luminosity.  This
particular projection highlights a weak feature of the FP that
otherwise would be completely masked, suggesting the existence of a
sort of warping (black curve in the figure; red in the electronic
version). Although just hinted in the bright part of the luminosity
function, this feature looks a bit more evident in its faint-end, which in
our sample is dominated by SDSS galaxies. To this concern, it is worth
noticing that this faint luminosity warp can hardly be attributed to a
possible upward bias of the SDSS low velocity dispersion measurements,
since, according to \citet{noao} (see also Figure~\ref{fig3}), such a
bias should in case work in the opposite direction. The different
shape of the NFPS and SDSS samples in this particular projection of
the FP explain the reasons why: (i) for the SDSS sample the
coefficient $a$ turns out to be always greater than in the case of the
NFPS sample; (ii) the faint-end luminosity cut influences the
coefficient $a$ of the FP more for the SDSS than for the NFPS
sample; (iii) the bright-end luminosity cut does not influence the FP
coefficients of the SDSS sample.

Although these analyses would benefit from a more robust statistic,
they lead us to suggest that the FP is likely a curved surface. This
fact has been recently claimed by \citet{desroches} and, in the
low-luminosity region, may actually indicate a first hint of the
connection between the FP of giant and dwarf ellipticals
(\citealp{nieto}, \citealp{held}, \citealp{peterson}). In
Section~\ref{sec5} we will also present a further hint of the
existence of the high luminosity warp of the FP suggested by
Figure~\ref{fig13}.

It is important to stress that the possible curvature of the FP may
give rise to different values of its coefficients when different
selection criteria, either chosen or induced by observations, are
acting to define galaxy samples. This fact represents a potentially
serious problem when the goal is to compare the tilt of the FP at low-
and high-redshifts, since it implies that a reliable comparison can be
done only if galaxy samples at quite different distances share the
same distributions of the photometric/kinematic properties, which is
indeed not usually the case.

Finally, we note that, according to the Chi-square values reported in the
right panels of Figure~\ref{fig7}, the scatter of the FP coefficients
is poorly consistent with the expected statistical uncertainties, even
after having reduced the annoying dichotomy between the NFPS and SDSS data
samples. This fact suggests that at least part of the observed scatter
must be somehow 'intrinsic' and resulting from a 'true' dependence of
the FP coefficients on the galaxy properties and/or on the local
environment and/or on the global cluster properties. The huge amount
of data available from the WINGS photometric catalogs allows us to
perform for the first time this kind of analysis.

\section{SYSTEMATICS OF THE FP COEFFICIENTS}\label{sec4}

In order to reduce the luminosity-driven bias of the FP coefficients
illustrated in the previous Section~\ref{sec32}, we decided to use in
this section only galaxies with $M_V<-19.5$ ($N_g$=1477). Even if this
luminosity cut-off does not remove completely the systematic
FP differences arising from the different sampling rules of the NFPS
and SDSS surveys (see Figure~\ref{fig11}), we guess it is able at
least to reduce them down to an acceptable level.

\subsection{FP versus galaxy properties and local environment}\label{sec41}

In Section~\ref{sec32} we have already shown that the FP coefficients
do depend on the average luminosity of the galaxies in the sample and,
therefore, on the average values of size and velocity dispersion (see
Figure~\ref{fig10}). These dependences concern the very shape of the FP
relation, since they involve the physical quantities defining the
relation itself. Now, besides these 'first order' dependences, we want
to check whether the FP relation varies with other galaxy properties
or the local environment. In particular, as far as the galaxy
properties are concerned, we test the ($B-V$) color, the Sersic index
Log($n$) and the axial ratio $b/a$, while the cluster-centric distance
$D_{CC}$ (normalized to $R_{200}$
\footnote{It is the radius at which the mean interior
overdensity is 200 times the critical density of the Universe.}) and
the local density Log($\rho$)\footnote{The local density around each
galaxy has been computed in the circular area containing the 10
nearest neighbors with $M_V<-19.5$: $\rho = 10/\pi R_{10}^2$ ($R_{10}$
in Mpc). The computation is a bit more complex for the objects close
to the edge of the WINGS CCD frames. A statistical background correction
of the counts has been applied using the recipe by \citealp{berta}} 
are used as test quantities of the local environment.

\begin{figure}
\vspace{-2.5cm}
\plotone{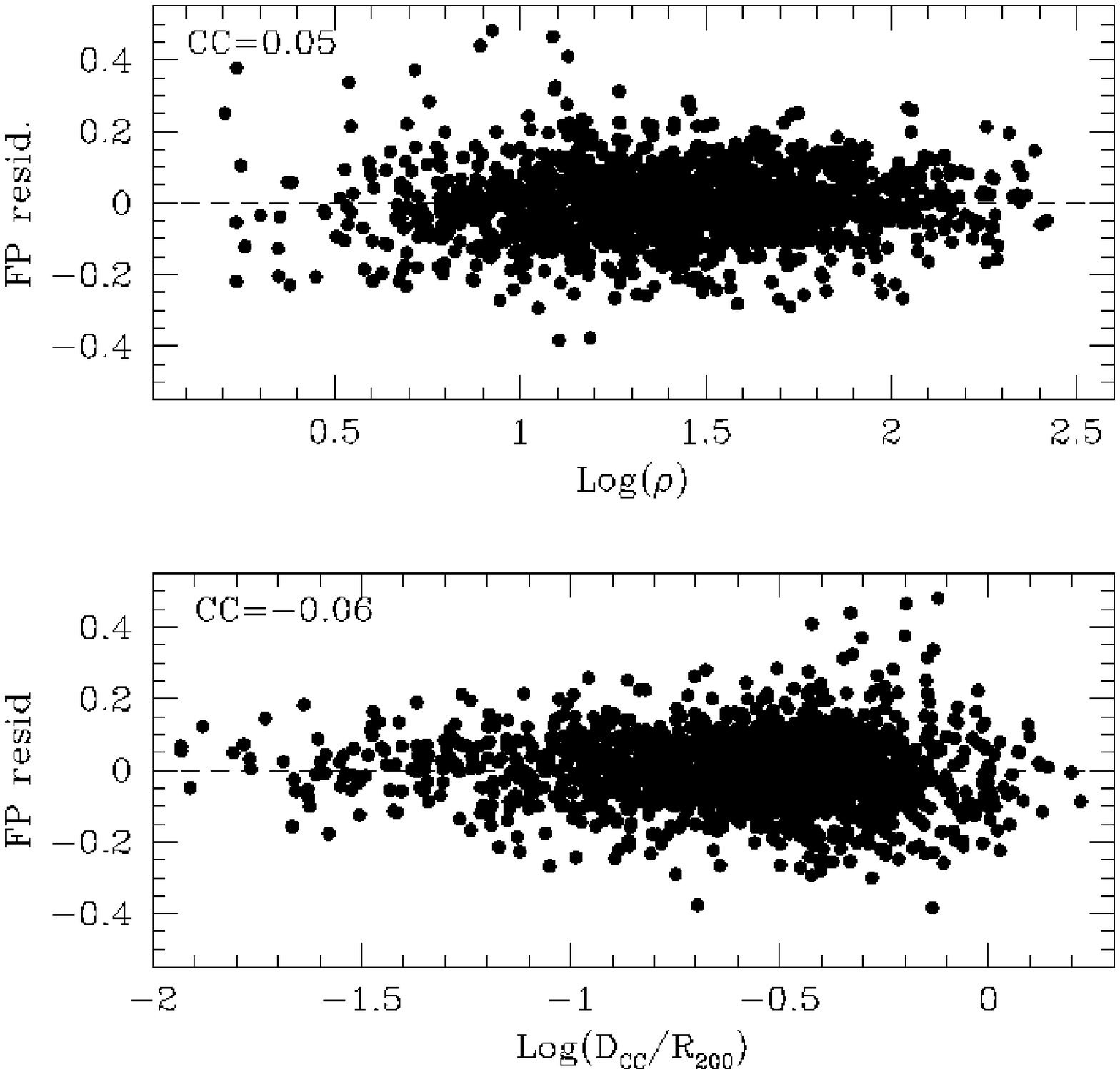}
\caption{
The residuals of the FP fit versus the local density (upper panel) and the
normalized cluster-centric distance (bottom panel). Note the lack of
correlation in this plots with respect to that found in Figure~\ref{fig15}.
\label{fig14}}
\end{figure}

A simple way to perform such kind of analysis is to correlate the test
quantities with the residuals of the FP relation obtained for the
global galaxy sample \citep{Jorg96}. For instance, in
Figure~\ref{fig14} the FP residuals are reported as a function of
both $D_{CC}$ and Log($\rho$). From this figure one would be led to
conclude that these two parameters do not influence at all the FP
coefficients. However, this method would be intrinsically unable to
detect any correlation if the barycentre of galaxies in the FP
parameter space does not change at varying the test quantity. In fact,
in this case, any change of the slope alone would produce a symmetric
distribution of the positive and negative residuals, keeping zero
their average value. For this reason, we preferred to perform the
analysis by evaluating the FP of galaxies in different bins of the
test quantities. Moreover, in order to get similar uncertainties of
the FP coefficients in the different bins, we decided to set free the
bin sizes, fixing the number of galaxies in each bin ($N_{bin}$).

\begin{figure}
\vspace{-2cm}
\includegraphics[angle=-90,scale=0.6]{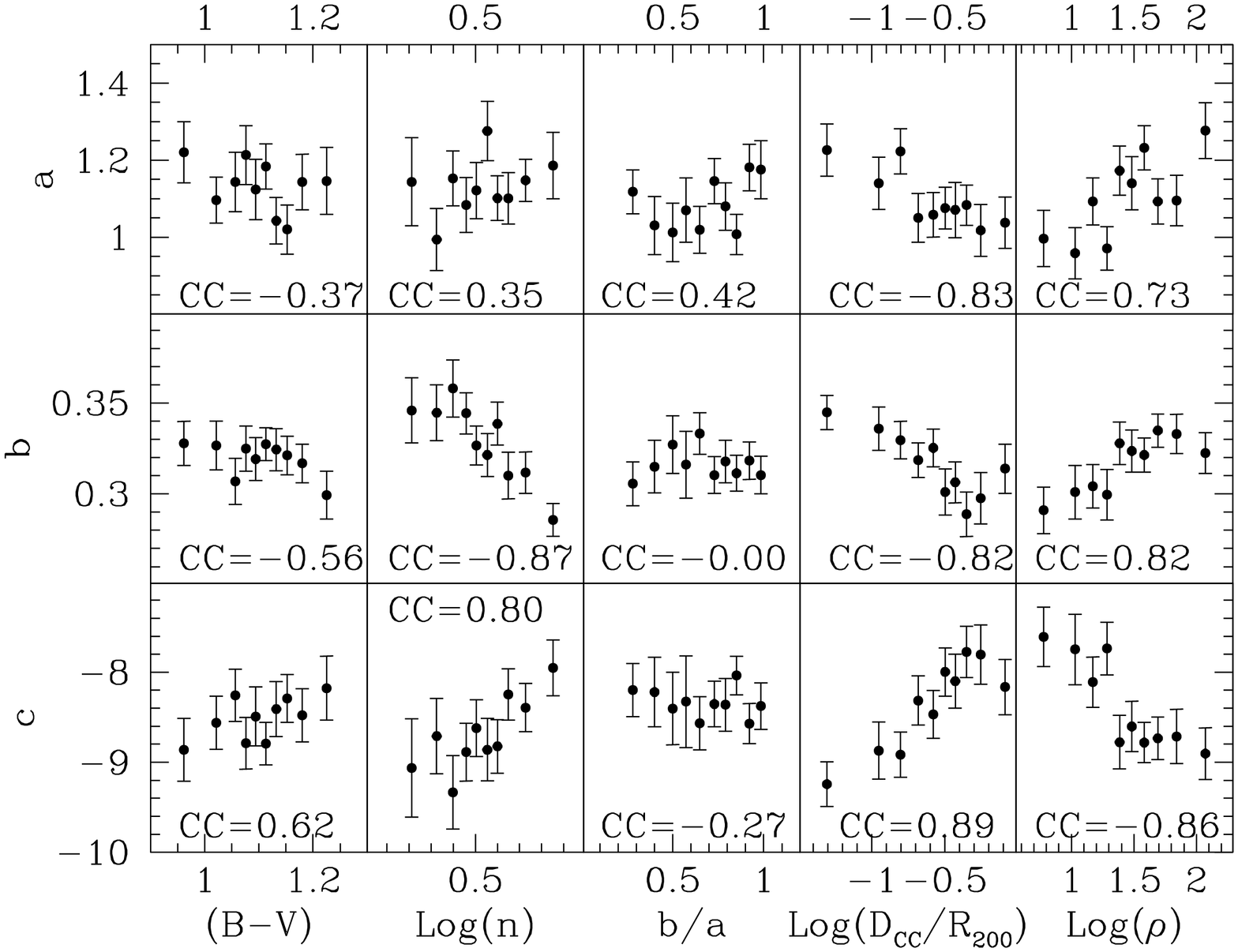}
\vspace{1cm}
\caption{
Average values of the the FP coefficients in different bins of color
[($B-V$)], Sersic index [Log($n$)], axial ratio [$b/a$], clustercentric
distance [Log($D_{CC}/R_{200}$)] and local density [Log($\rho$)] as
a function of the median values of the same quantities inside the
bins. The number of galaxies per bin is fixed to 150 and the BCGs are
assumed to coincide with the centers of the clusters. The correlation
coefficients of the different pairs of bin-averaged quantities are
also reported in the panels. Note the strong correlations 
between the FP coefficients and the environment parameters $D_{CC}$
and $\rho$.
\label{fig15}}
\end{figure}

In Figure~\ref{fig15} the average values of the the FP coefficients in
different bins of the test quantities are plotted as a function of the
median values of the quantities themselves inside the bins. The panels
also report the correlation coefficients ($CC$) of the different pairs
of bin-averaged quantities. In these plots we set $N_{bin}$=150 and
assumed the centers of the clusters to coincide with the position of
the BCGs. However, the trends and the correlation coefficients in the
figure remain almost unchanged if we set (for instance) $N_{bin}$=200
and assume that the cluster centers coincide with the maximum of the
X-Ray emission.

It is worth stressing that the FPs we obtain with the outlined
procedure for each bin of the test quantities do not refer to real
clusters. They are actually relative to ideal samples for which some
galaxy/environment property is almost constant (for instance: constant
local density).

\begin{figure}
\vspace{-1cm}
\epsscale{.90}
\plotone{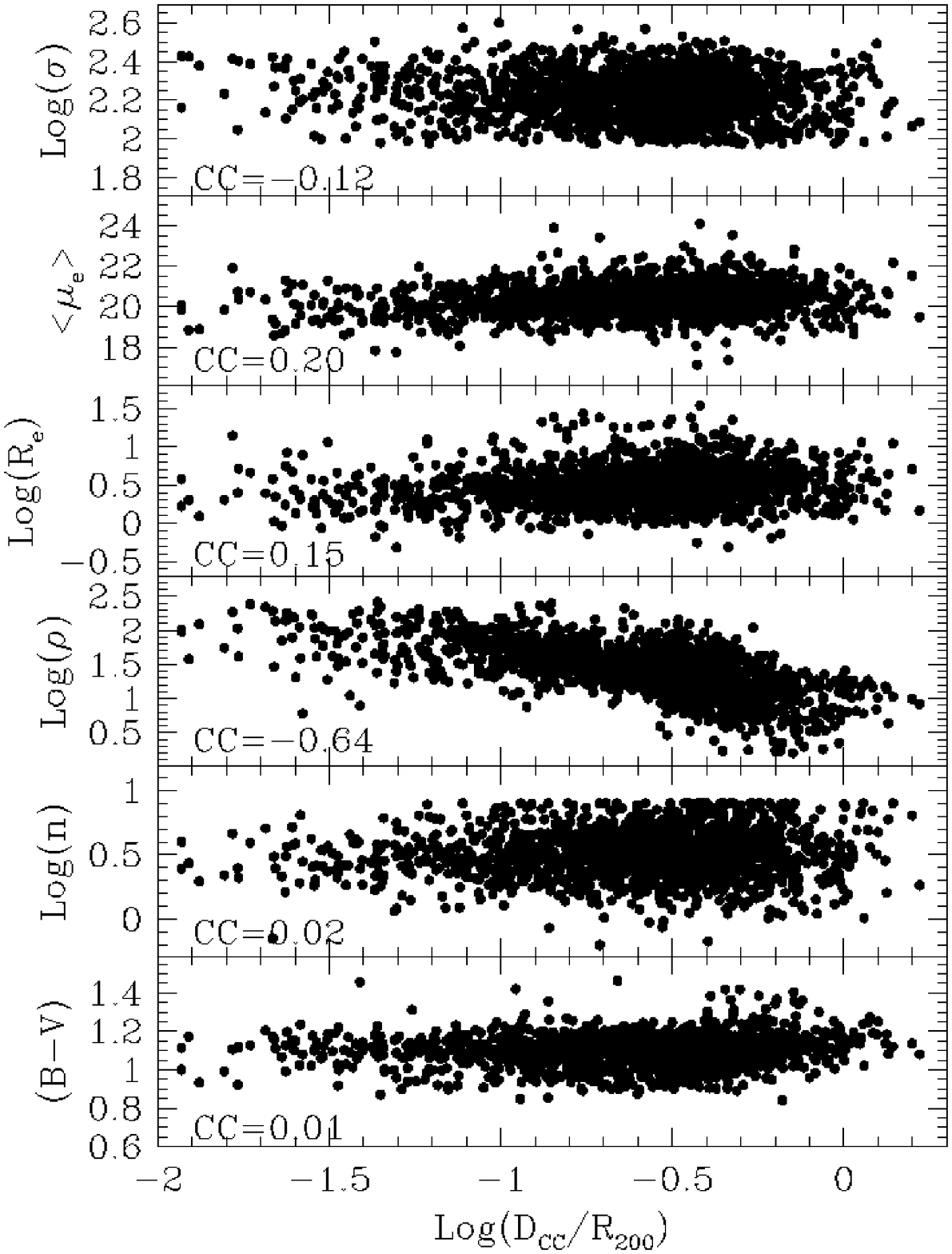}
\vspace{-1.0cm}
\caption{ Normalized cluster-centric distance versus velocity
dispersions, effective surface brightness, effective radii, local
density, Sersic index and color for our sample of early-type galaxies
with $M_V<-19.5$. Note the well known (obvious) dependence of the
local density on the clustercentric distance, while the other panels 
display the substantial lack of correlations with the other variables. 
The obvious cut at large $Log(n)$ is due to the intrinsic limit of 
GASPHOT to give Sersic index $n>8$.
\label{fig16}}
\end{figure}

At variance with the conclusions one could draw from Figure~\ref{fig14},
it is clearly show in Figure~\ref{fig15} that strong correlations exist
between the FP coefficients and the environment parameters ($D_{CC}$
and $\rho$), while the correlations are less marked (or absent) with
the galaxy properties. We have verified that the average (and median)
absolute magnitudes do not vary significantly in the different bins of
the test quantities Log($D_{CC}$) and Log($\rho$). Thus, the
correlations among these quantities and the FP coefficients cannot be
induced by the above mentioned dependence of the FP coefficients on
the absolute magnitude (see Sec.~\ref{sec32}). Moreover, the lack of
correlation in the three uppermost panels of Figure~\ref{fig16} 
rules out the possibility that the above trends just reflect similar
trends involving the very physical quantities that define the FP.

Figure~\ref{fig15} suggests that, in the FP, the dependences on both
velocity dispersion and average surface brightness of galaxies become
lower and lower as the distance from the cluster center increases.
Looking at the two leftmost panels of Figure~\ref{fig15}, one could
wonder if the further dependences of the FP coefficients on both 
the Sersic index (stronger) and the color (weaker) are merely reflecting
the correlation with the clustercentric distance. The
lack of correlation in the two lowest panels of Figure~\ref{fig16}
help to clarify this point, suggesting that light concentration and
color could actually be additional (independent) physical ingredients
of the FP recipe. It is also interesting to note in Figure~\ref{fig15}
that the $b$ coefficient correlates quite well with the Sersic index
$n$, while $a$ does not. This is likely because $b$ is the coefficient
associated with the photometric parameter \muem\ , which is in turn
obviously related to the concentration index $n$. Finally, we note
that, from the very (linear) expression of the FP, most of the
dependences of the $c$ coefficient on the various test quantities in
Figure~\ref{fig15} are likely induced by the corresponding dependences
of the $a$ and $b$ coefficients, any increase of the last ones
producing necessarily a decrease of the former one, and viceversa.

The trends observed in Figure~\ref{fig15} further confirm that the
FP coefficients depend on the particular criteria used in selecting
the galaxy sample. They are also likely able to explain the large
scatter of the FP coefficients which is found even after removal of
the luminosity-driven bias discussed in Section~\ref{sec32} (high
values of $\chi^2_{\nu}$ and $P_{\nu}$ in Figure~\ref{fig7}).

\subsection{FP versus global cluster properties}\label{sec42}

\begin{figure}
\vspace{-1cm}
\plotone{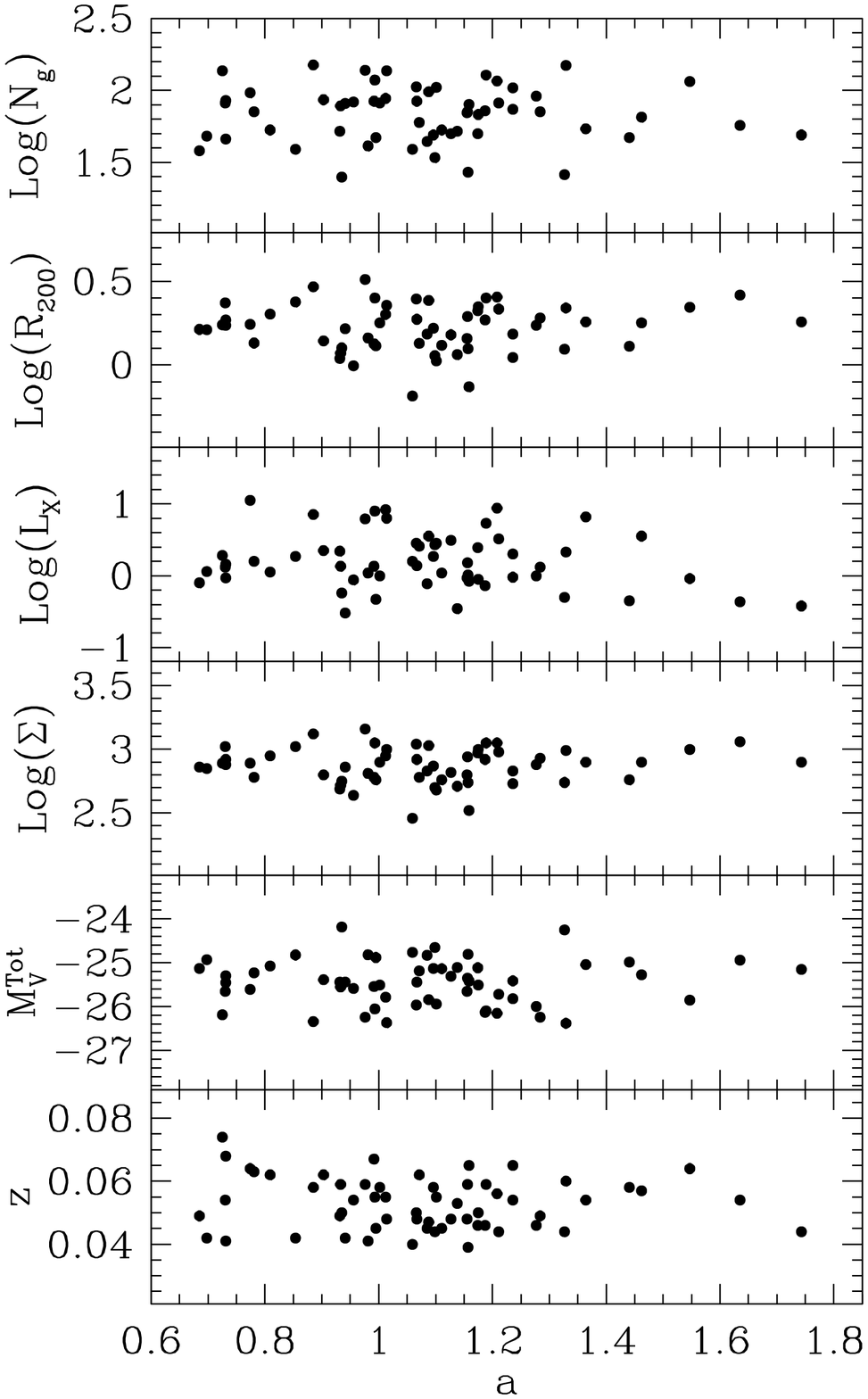}
\vspace{-1cm}
\caption{
From top to bottom: the FP coefficient $a$ versus the number of
galaxies (with $M_V<-19.5$) in the cluster, the radius of the cluster $R_{200}$, the
X-ray luminosity, the $rms$ of peculiar velocities of galaxies in the cluster, the
integrated V-band luminosity (again with $M_V<-19.5$) and the
redshift. No significant correlations are found.
\label{fig17}}
\end{figure}

We have also explored the possible dependence of the FP coefficients
(in particular of the coefficient $a$) on several measured quantities
related to the global cluster properties. Tentative correlations have
been performed with the velocity dispersion of the galaxies in the
clusters, with the X-Ray luminosity, with redshift, with the
integrated V-band luminosity (within $M_V=-19.5$) of the clusters,
with different kinds of cluster radii, with the average Sersic index
of the cluster galaxies, with the average Log($M/L$), etc. Some of
these plots are shown in Figure~\ref{fig17}. No significant
correlations have been found.

\subsection{Can we provide a general recipe for deriving the FP ?}\label{sec43}

From the analysis performed in Section~\ref{sec3}, the differences in
the FP coefficients appear to be related to sampling aspects (i.e. the
luminosity cut-off). In Section~\ref{sec4} we have shown that, although
not depending on the global cluster properties, the FP coefficients
are also strongly related to the environmental properties of galaxies 
($D_{CC}$ and $\rho$) and to their internal structure (Sersic index). 
These dependences likely concern the very formation history
of galaxies and clusters. They are not strictly referable as sampling
effects, but we can of course always speak of sampling, as far as they
translate into the photometric properties of galaxies. This let us
understand that the various dependences are actually linked
each other and it is not easy to isolate each of them. Moreover, it
is worth stressing that the previous analyses (never tried before)
have been made possible just because we have at our disposal a huge
sample of galaxies, obtained putting altogether data from many
different clusters. When dealing with the determination of the FP for
individual (possibly far) clusters, we usually must settle for what we
actually have, that are a few galaxies (a few dozens, in the most
favourable cases) with different luminosities and structures, located
in a great variety of environments. In this cases, we can hardly renounce
to each single galaxy and the above dependences turn out to be irreparably 
entangled each one another.

From the previous remarks it stands to reason that, even with our
large sample of galaxies, to provide a general recipe for determining
unbiased coefficients of the FP in individual clusters is far from
being a realistic objective. The best we can do is to remove from our
global (W+N+S) sample the low-luminosity galaxies
($M_V>$-19.5; see Section~\ref{sec32}) and to provide, for
this restricted sample, the FP coefficients obtained with both MIST and
ORTH fitting algorithms. They are:

$a=1.097\pm 0.020; b=0.318\pm 0.004; c=-8.41\pm 0.097$ (MIST)

$a=1.208\pm 0.052; b=0.318\pm 0.010; c=-8.65\pm 0.19$, (ORTH)

\noindent
which we assume to define the global (unbiased, as far as possible),
V-band FP of early-type galaxies in nearby clusters. The MIST
coefficients are more suitable for distance determination, while
the ORTH ones more properly define the physical relation among the
quantities involved in the FP.

Columns 10-15 of Table~\ref{tbl3} report the FP coefficients of the
individual clusters obtained running MIST just for galaxies with
$M_V<$-19.5.

\section{THE $M/L_V$ RATIO OF EARLY-TYPE GALAXIES IN NEARBY CLUSTERS}\label{sec5}

The variation of the mass-to-light ratio with luminosity is the most
popular explanation for the tilt of the FP with respect to the virial
expectation. Therefore, it is important to analyse the $M/L$ ratio of
cluster galaxies with our extensive photometric data, which account
for the non homologous structure of the ETGs by means of the Sersic
parameter $n$.

According to \citet{Sperello} (see also \citealp{Michard80}), we
calculate the dynamical mass of galaxies using the formula:
$M/M_{\odot}=K_V(n)\sigma^2R_e/G$, where the virial coefficient
$K_V(n)$ is a decreasing function of the Sersic index $n$
(\citealp{Bertin02}) and $G$ is the gravitational constant.

\begin{figure}
\plotone{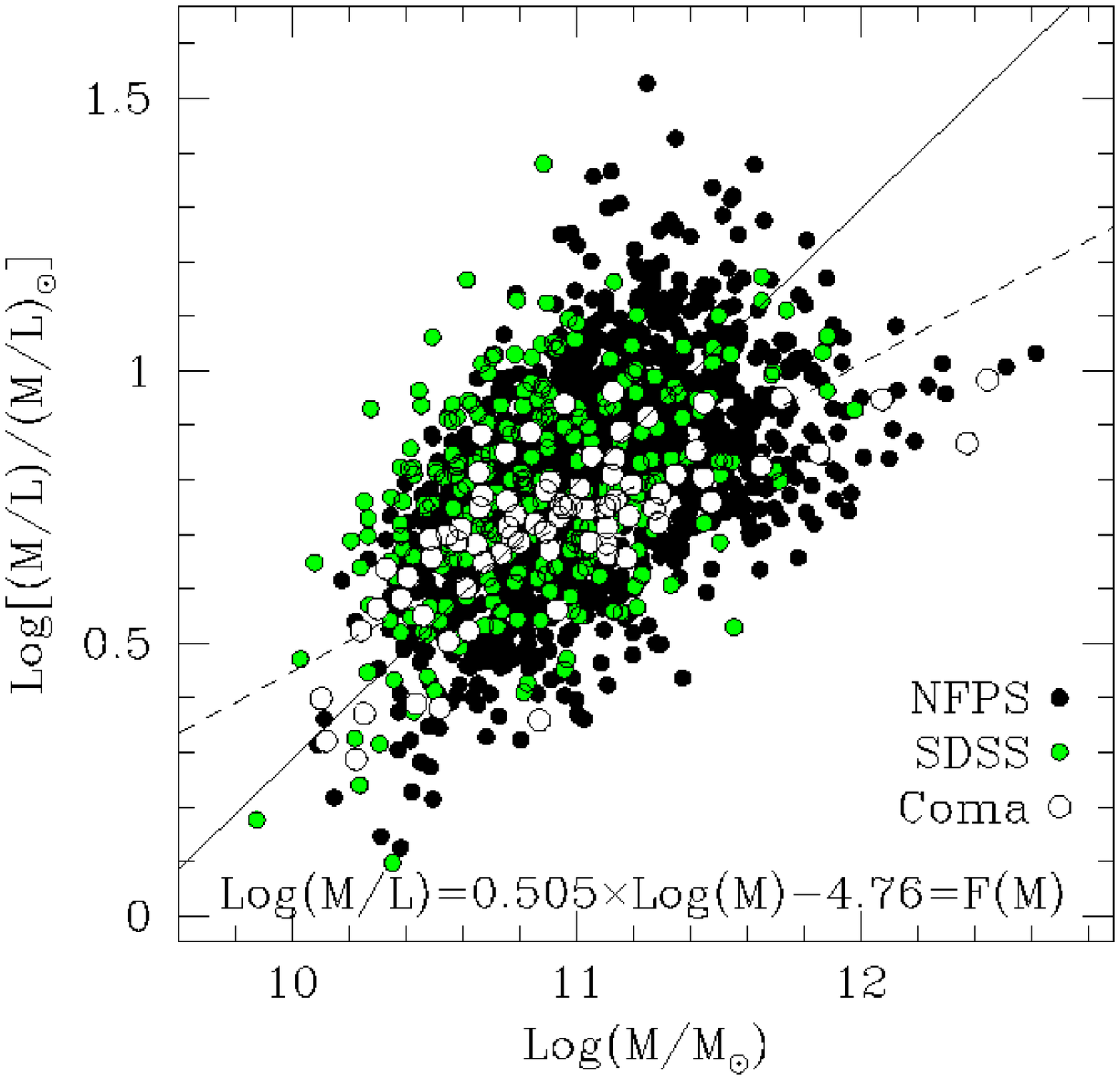}
\caption{
Mass-to-light ratio versus dynamical mass for our global galaxy
sample. Black and gray (green in the electronic version) dots refer 
to W+N and W+S galaxies, respectively, while open
dots refer to a sample of galaxies in Coma (see text). The solid line
gives the orthogonal fit of the W+N+S data, while the dashed line
reports the fit for the Coma sample.
\label{fig18}}
\end{figure}

Figure~\ref{fig18} shows the mass-to-light ratio as a function of mass
for our global galaxy sample. The full straight line in the figure
represents the linear fit obtained minimizing the weighted sum of
perperdicular distances from the line itself [see the equation F(M) in
the figure]. The open dots refer to the sample of galaxies in Coma
given by \citet{Jorg96}, with the relative orthogonal fit represented
by the dashed line. Note in particular that in our global sample
the scatter of the residuals relative to the best-fit relation is greater 
than in the Coma sample (0.19 vs. 0.11; for the NFPS and SDSS samples
the scatters are 0.18 and 0.21, respectively). However,
we recall that \citet{Jorg96} derived the photometric parameters (\re\
and \muem\ ) and the mass by assuming \rq\ luminosity profiles, while
we adopted the more general Sersic profiles. This might also explain 
the fact that the slope of the relation for our global sample [0.511($\pm$0.019)]
is much larger that in the Coma sample [0.28($\pm$0.028)]. By the way,
the slopes we found for the NFPS and SDSS samples separately, are quite consistent
each other, within the errors [0.522($\pm$0.022) and 0.600($\pm$0.052),
respectively].

\begin{figure}
\vspace{-4cm}
\hspace{-2cm}
\includegraphics[angle=-90,scale=0.75]{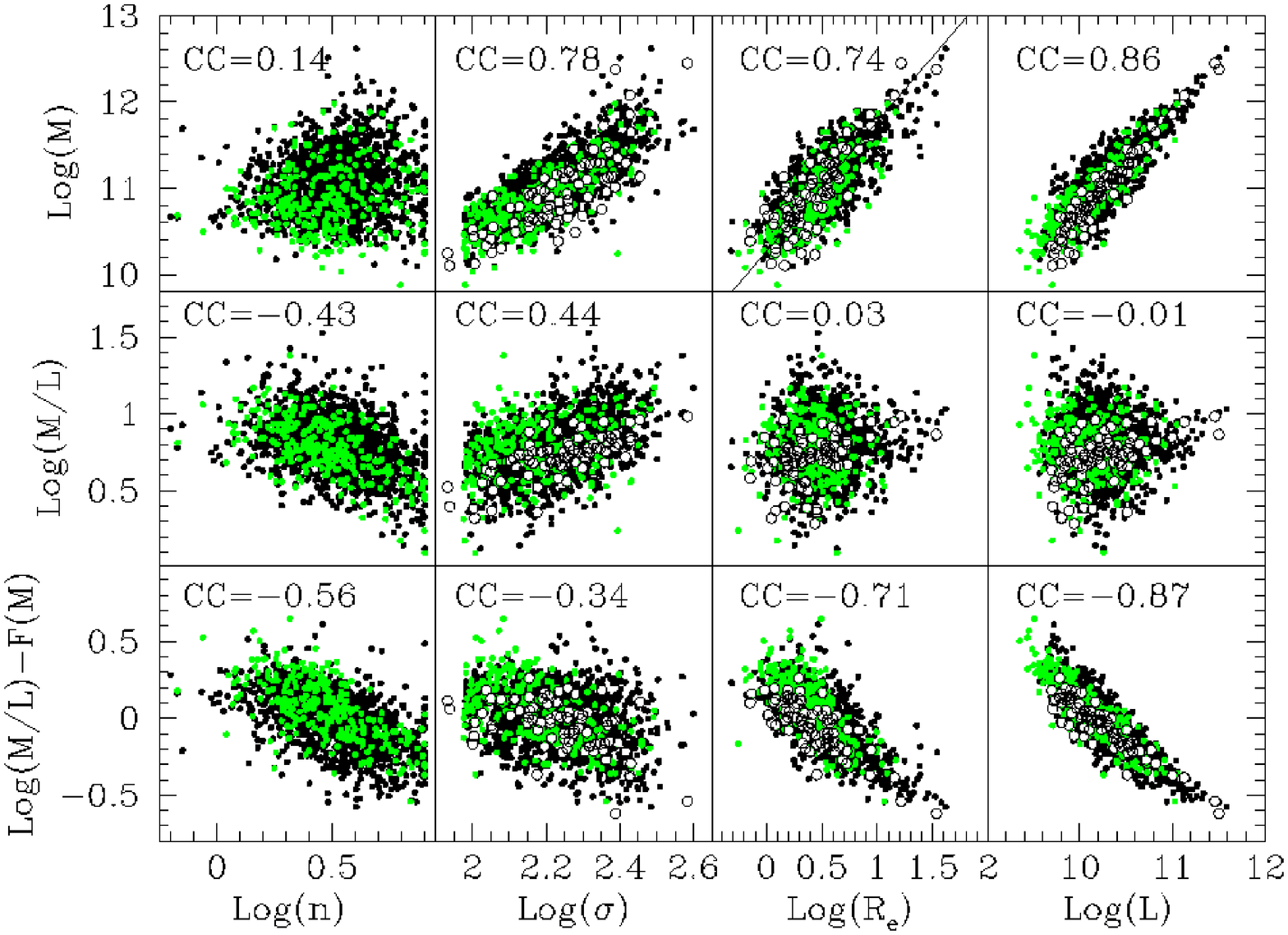}
\caption{
Correlations among measured (abscissa) and evaluated (ordinate) 
quantities involved in the mass-to-light ratio estimate. In each
panel the proper correlation coefficient is also reported. Symbols
are as in Figure~\ref{fig18}. The straight line reported in the plot
Log(M) - Log$R_e$ represents the orthogonal best-fit of the data
we discuss in the last paragraph of this Section.
\label{fig19}}
\end{figure}

Figure~\ref{fig19} reports several plots showing the correlations
among different measured and evaluated quantities involving the
mass-to-light ratio estimate. In particular we test (at the ordinate)
dynamical masses, mass-to-light ratios and residuals [Log(M/L)-F(M)]
of the relation in Figure~\ref{fig18} versus (at the abscissa) Sersic
indices, velocity dispersions, effective radii and luminosities.

Some of the correlations in Figure~\ref{fig19} are well known
(i.e. mass vs. luminosity), obvious (i.e. $M/L$ residuals
vs. luminosity), or expected by definition (i.e. mass vs. $\sigma$ and
\re\ ). Less obvious seem to be some other correlations (i.e. $M/L$
vs. $\sigma$ and $M/L$ residuals vs. $\sigma$ and \re\ ) or lack of
correlations (i.e. $M$ vs. Sersic index, $M/L$ vs. \re\ and
luminosity). For instance, according to the formula we used to derive
the dynamical mass, it should be a strongly decreasing function of the
Sersic index (see \citealp{Bertin02}), while the correlation
coefficient of the plot $M$-$n$ in Figure~\ref{fig19} is slightly
positive. Moreover, in the same figure the $M/L$ ratio does not seem
to correlate at all with either radius or luminosity, while a
correlation $M/L$-$L$ has been often claimed to explain the 'tilt' of
the FP.

In Figure~\ref{fig19} we find of particular interest the correlation
between $M/L$ and Sersic index and that between $M/L$ residuals and
Sersic index. The first one, coupled with the lack of correlation
between $M/L$ and luminosity (which is indeed expected for non
homologous ETGs, as suggested by \citealp{Trujillo}), indicates that,
for a given luminosity, the galaxies showing lower light concentration
(lower Sersic index) are more massive (more dark matter?).

The second, even stronger correlation is quite interesting as well. In
fact, from the very definition of the $M/L$ residuals of the relation
in Figure~\ref{fig18}, for a given dynamical mass, the lower the
residual, the brighter the galaxy. Therefore, the correlation in
Figure~\ref{fig19} between $M/L$ residuals and Sersic indices implies
that (again for a given mass) the higher the light concentration
(Sersic index), the brighter the galaxy.

Thus, the picture emerging about the influence of the light
concentration in determining dynamical mass and luminosity of ETGs is
that: (i) for a given luminosity, the higher the light concentration,
the lower the dynamical mass; (ii) for a given dynamical mass, the
higher the light concentration, the higher the luminosity. This
twofold dependence on the Sersic concentration index is expressed by
the linear equation:

Log(n)=1.60$\times$Log(L)$-$1.16$\times$Log(M)$-$2.93,

we have derived minimizing the orthogonal distances from the fitting 
plane of the points in the parameter space ($n$,$L$,$M$). Note
that the correlation coefficient between the Sersic index computed
from this equation and the measured one is $CC$=0.59.

Still concerning the influence of the light concentration on the
mass-to-light ratio of early-type galaxies, it is well known that
the Sersic index $n$ correlates with the velocity dispersion
(\citealp{Graham}). Thus, it is not meaningless wondering if the
correlations involving $n$ in Figure~\ref{fig19} just reflect the
correlations with $\sigma$. The upper and middle panels of the figure 
clearly rule out this hypothesis as far as the correlations with mass 
and mass-to-light ratio are concerned (both are positive for $\sigma$, 
while for $n$ they are close to zero and negative, respectively).
Instead, the correlations of the $M/L$ residuals with $n$ and $\sigma$ 
(lower panels) have the same sign. It is worth noting, however,
that the correlation turns out to be tighter with $n$ than
with $\sigma$ and that the same happens (even if with opposite trends)
also for the $M/L$ ratio in the middle panels of the figure.
This might suggest that the driving parameter for $M/L$ is actually the
light concentration and that the trends with $\sigma$ are just
consequence of that.

We have previously guessed that the different slopes we find in the
relation ($M/L$--$M$) between our sample and the Coma sample could be
at least partially due to the different models of luminosity profiles
used to derive the photometric parameters of galaxies (Sersic law and
$r^{1/4}$ law, respectively). Now, we could legitimately guess that
the correlations shown in the leftmost panels of Figure~\ref{fig19}
are artificially produced by the use of the Sersic law in deriving the
parameters $R_e$ and $K_V$, involved in the computation of the galaxy
mass. Actually, $K_V$ turns out to be a decreasing function of the
Sersic index (see \citealp{Bertin02}), just like the $M/L$ ratio and
the $M/L$ residuals in Figure~\ref{fig19} (but, in the same figure
note the direct, although weak, correlation between the mass and the
Sersic index!).

\begin{figure}
\plotone{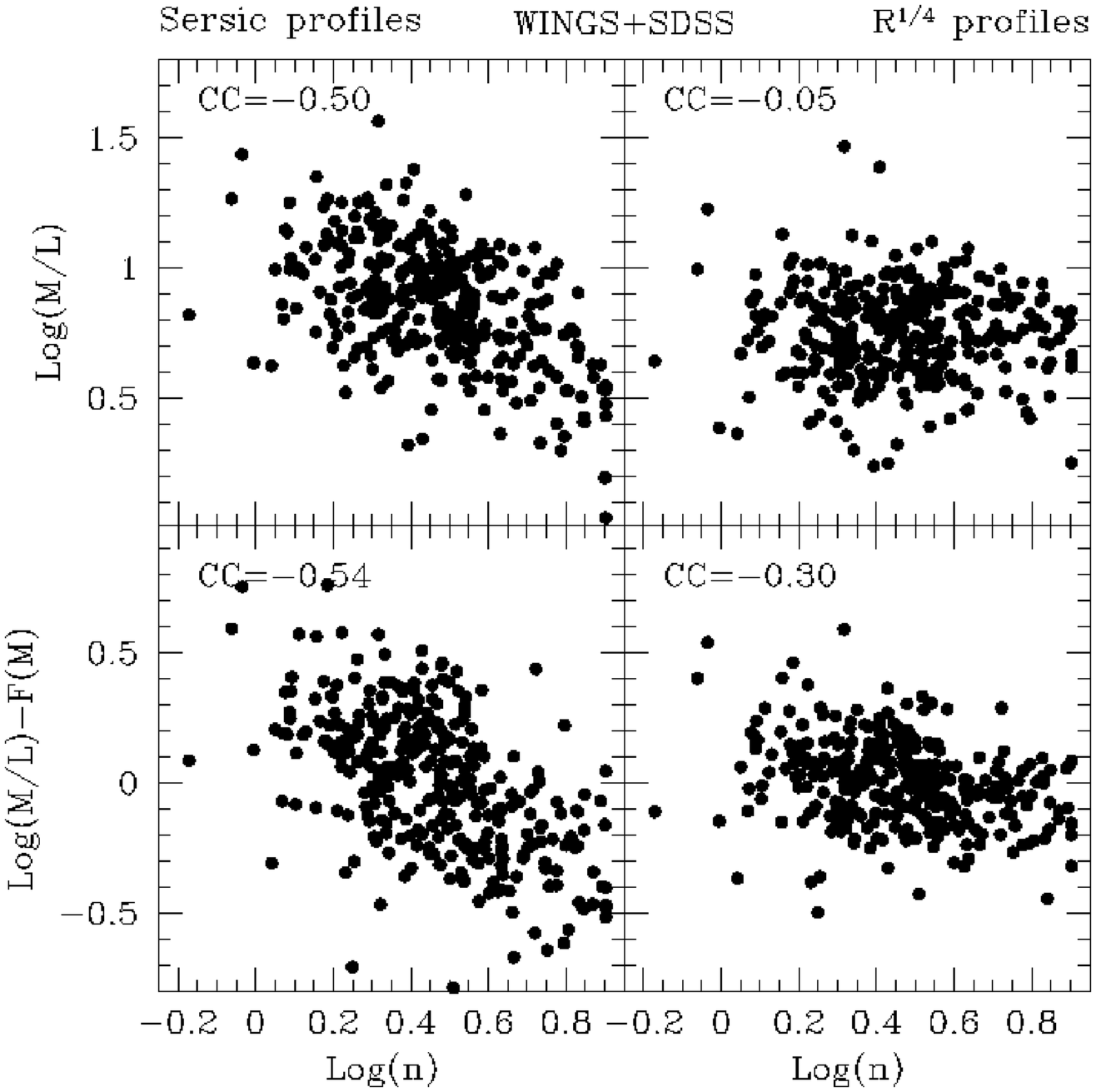}
\caption{
The $M/L$ ratio and the residuals of the $M/L$--$M$ relation versus
the Sersic index for the W+S galaxy sample alone. Left and right
panels illustrates the relations obtained using the Sersic and
$R^{1/4}$ surface photometry parameters, respectively, in the
computation of masses and luminosities. Again, the correlation
coefficients are reported in each panel.
\label{fig20}}
\end{figure}

Trying to clarify these points, we have re-calculated the masses and
the luminosities of the early-type galaxies in the original W+S sample 
(397 objects, before selection on $\sigma$ and $M_V$) using
the surface photometry parameters provided by the SDSS database
($r^{1/4}$ profiles). In Figure~\ref{fig20} we plot the $M/L$ ratio and
the residuals of the $M/L$--$M$ relation versus the Sersic index for
the W+S galaxy sample alone. Left and right panels illustrate the
relations obtained when masses and luminosities are computed using the 
Sersic and $r^{1/4}$ surface photometry
parameters, respectively. In the former case, the correlation coefficients turn
out to be undistinguishable from those obtained for the whole (W+N+S)
galaxy sample (see Figure~\ref{fig19}). In the latter case the
correlations are less pronounced, but still they are in place, as indicated by
10,000 random extractions of couples of uncorrelated vectors having the
same dimension (397) and distributions of the real ones. In fact, the
probability that the correlation coefficients of the real sample in the
right panels of Figure~\ref{fig20} are
drawn from a parent population of uncorrelated quantities turns out to
be very small: $\sim$0.005 and $\sim$0 for the correlations in the
upper-right and lower-right panels,
respectively. This enforces our previous conclusions about the
dependence of masses and luminosities of early-type galaxies on the
Sersic index. The weaker correlations found with the $r^{1/4}$
profiles if compared with the Sersic profiles, are likely the
consequence of having forced the real luminosity structure of galaxies
to obey the de~Vaucouleurs law. Finally, we mention that the slope of
the relation ($M/L$--$M$) for the W+S sample turns out to be 0.47 and
0.38 with the Sersic-- and $r^{1/4}$--law approaches, respectively,
thus confirming our previous guess that it is influenced by the
assumption about the luminosity profile of galaxies (see the
comparison between our sample and the Coma sample in
Figure~\ref{fig18}).

In a recent paper \citet{Robertson} claim that the tilt of the FP is
closely linked to dissipation effects during galaxy formation. They
show the results of their simulations in a plot of $M/M_{\odot}$ vs.
the radius ($R$) of the galaxies. Figure~\ref{fig19} also shows a
similar plot for our galaxy sample. Although a correlation between $M$
and $R_e$ is expected from the very definition of dynamical mass, we
note in this plot that the more massive objects are preferably found
below the orthogonal best-fit of the data distribution (by the way,
these objects are those deviating in the high luminosity region
from the FP projection in Figure~\ref{fig13}). High luminosity objects 
also deviate with respect to the bulk of the early-type population in 
the $R - L$ relation, that for our global sample has the same slope 
found by \citet{Bern07} ($R\propto L^{0.68}$). The systematically 
larger size of these galaxies in that relation may be consistent with 
the results of the simulations by \citet{Robertson}, if one invokes 
the dry dissipationless merger mechanism for their formation. 
It again points towards the hypothesis that the FP relation might be non
linear, this time in its high-mass region (see Section~\ref{sec32} for
a similar finding in the low-mass region). Thus, in the parameter
space of the FP the real surface defined by ETGs could be slightly
bent, reflecting the different formation mechanisms producing the
present day ETGs. In Section~\ref{sec32} we have seen that
part of the scatter of the FP coefficients is just due to
such an effect, coupled with the different statistical properties of
the galaxy samples. As a consequence, to draw any conclusion about
luminosity evolution and downsizing effect might be dangerous (the
slopes of the relations depend on the sample selection rules/biases),
unless the galaxy samples involved in these analyses, even spanning
wide ranges of redshift, share the same distributions of
photometric/kinematic properties of galaxies.

\section{CONCLUSIONS}\label{sec6}

We have derived the Fundamental Plane of early-type galaxies in 59
nearby clusters ($0.04<z<0.07$) by exploiting the data coming from
three big surveys: WINGS(W), to derive \re\ and \muem\ , and
NFPS+SDSS(N+S), to derive $\sigma$. The fits of the FP, obtained for
the global samples W+N and W+S, as well as for each cluster, have
revealed that the FP coefficients span considerable intervals. 
By means of extensive simulations, we have demonstrated
that this spread is just marginally consistent with the statistical noise due to
the limited number of galaxies in each cluster. It seems at least
partly due to a luminosity-driven bias depending on the statistical
properties of the galaxy samples. These can be induced both by
observing limitations and by selection rules. In fact, even if the
best-fitting solution obtained for the global W+N+S dataset does not
differ significantly from previous determinations in the literature,
systematic different FP coefficients are found when the galaxy samples
are truncated in the faint-end part at different cut-off absolute magnitudes. We speculate
that, rather than a plane, the so called FP is actually a curved surface,
which is approximated by different planes depending on the different regions
of the FP space occupied by the galaxy samples under analysis. To this
concern, we could go farther on in the speculation, suggesting that a
bent FP could be, at least partially, reconciled with the numerical 
simulations in $\Lambda$CDM cosmology (see \citealp{Borriello}). 
By the way, such a speculation could also be supported by the large 
scatter we find in the $M/L$--$M$ relation, at variance with other 
determinations, whose tightness has been sometime invoked to rule 
out the hierarchical scenario.

Perhaps the most interesting result of the present analysis concerns the
dependence of the FP coefficients on the local environment, which
clearly emerges when we derive the FP in different bins of the
cluster-centric distance and local density. Finally, we do not find
any dependence of the FP coefficients on the global properties of 
clusters.

Concerning the $M/L$ ratio, we also find that both $M/L$ and the
residuals of the $M/L$--$M$ relation turn out to be anti-correlated with
the Sersic indices. These trends could imply that, for a given
luminosity, more massive galaxies display a lower light concentration,
while for a given dynamical mass, the higher the light concentration,
the brighter the galaxy.

The main results of this work can be summarized as follows:
\begin{itemize}
\item the FP coefficients depend on the adopted fitting technique and
(marginally) on the methods used to derive the photometric parameters \re\ and \muem\ ;
\item The observed scatter in the FP coefficients cannot be entirely
ascribed to the uncertainties due to the small number statistics;
\item the FP coefficients depend on the distributions of photometric/kinematic properties 
of the galaxies in the samples (mainly on the faint-end luminosity cut-off);
\item the FP coefficients are strongly correlated with the environment
(cluster-centric distance and local density), while the correlations
are less marked (or absent) with the galaxy properties (Sersic index,
color and flattening);
\item the FP coefficients do not correlate with the global properties
of clusters (radius, velocity dispersion, X-ray emission, etc..);
\item the distribution of galaxies in the FP parameter space suggest that the
variables \re, \muem\ , and $\sigma$ define a slightly warped surface.
Forcing this surface to be locally a plane causes a systematic variation the FP 
coefficients, depending on the selection rules used to define the galaxy sample;
\item using the FP as a tool to derive the luminosity/size evolution of ETGs 
may be dangerous, unless the galaxy samples involved in the analysis are
highly homogeneous in their average photometric properties;
\item the $M/L$ ratio is not correlated with $L$ when the non homology
of ETGs is taken into account. This is an indication that most of
the tilt of the FP is indeed due to dynamical and structural
non-homology of ETGs;
\item the mutual correlations among mass, luminosity and light concentration
of ETGs indicate that, for a given mass, the greater the light
concentration the higher the luminosity, while, for a given
luminosity, the lower the light concentration, the greater the mass;
\item the bending of the FP and the large scatter found in the $M/L$--$M$ relation 
could, at least partially, reconcile the FP phenomenology with the hierachical merging
scenario of galaxy formation.
\end{itemize} 

By exploiting the galaxy mass estimates coming from both the K-band
WINGS data and the spectro-photometric analysis of the galaxies in the
WINGS survey (\citealp{Fritz}), in a following paper we will go into
more depth about the scaling relations involving mass, structure and
morphology of galaxies in nearby clusters.

\acknowledgments
{We wish to thank the anonymous referee, since her/his useful comments helped us 
to improve the final version of the paper. We also wish to thank our colleagues
from the Astronomical Observatory of Napoli (La Barbera et al. 2000) for having 
kindly provided us with the fitting tool MIST, which we have intensively used
in this paper to derive the FP coefficients.}


\begin{deluxetable}{lccccccccc}
\tabletypesize{\small}
\tablewidth{0pt}
\tablecaption{The Cluster sample.\label{tbl1}}
\tablehead{
\colhead{Cluster} & \colhead{z(NED)} & \colhead{$\Sigma$} & \colhead{$L_X$} & \colhead{$M_{tot}$} & \colhead{$R_{200}$} & \colhead{$M_V(BCG)$} & \colhead{$N_{W+N}$} & \colhead{$N_{W+S}$} & \colhead{$N_{W+[N\&S]}$}}
\startdata
       A0085  &    0.0551     &     1152    &     44.92    &      -25.75    &     2.590    &     -23.80    &   41  &   46  &  23  \\
        A119  &    0.0442     &      951    &     44.51    &      -25.74    &     2.149    &     -23.70	   &   46  &   15  &  13  \\
        A133  &    0.0566     &      823    &     44.55    &      -25.28    &     1.849    &     -23.27	   &   23  &   --  &  --  \\
        A160  &    0.0447     &      806    &     43.58    &      -25.13    &     1.822    &     -22.89	   &   17  &   18  &  12  \\
        A168  &    0.0450     &      613    &     44.04    &      -25.15    &     1.385    &     -22.85	   &   --  &   12  &  --  \\
        A376  &    0.0484     &      906    &     44.14    &      -25.45    &     2.044    &     -23.15    &   27  &   --  &  --  \\
       A548b  &    0.0416     &      928    &     43.48    &      -25.46    &     2.099    &     -22.96	   &   22  &   --  &  --  \\
        A602  &    0.0619     &      754    &     44.05    &      -24.97    &     1.691    &     -22.52	   &   14  &   13  &   9  \\
        A671  &    0.0502     &      938    &     43.95    &      -25.47    &     2.114    &     -23.58	   &   --  &   16  &  --  \\
        A754  &    0.0542     &     1101    &      44.9    &      -26.02    &     2.476    &     -23.67	   &   46  &   --  &  --  \\
        A780  &    0.0539     &      751    &     44.82    &      -25.07    &     1.689    &     -23.31	   &   16  &   --  &  --  \\
       A957x  &    0.0460     &      710    &     43.89    &      -24.94    &     1.604    &     -23.44	   &   17  &   22  &   9  \\
        A970  &    0.0587     &      865    &     44.18    &      -25.20    &     1.941    &     -22.31	   &   25  &   --  &  --  \\
       A1069  &    0.0650     &      723    &     43.98    &      -25.39    &     1.618    &     -23.22	   &   20  &   --  &  --  \\
       A1291  &    0.0527     &      479    &     43.64    &      -24.88    &     1.079    &     -22.41	   &   --  &   13  &  --  \\
      A1631a  &    0.0462     &      803    &     43.86    &      -25.71    &     1.813    &     -22.93	   &   22  &   --  &  --  \\
       A1644  &    0.0473     &     1092    &     44.55    &      -25.89    &     2.465    &     -23.72	   &   41  &   --  &  --  \\
       A1668  &    0.0634     &      668    &      44.2    &      -25.32    &     1.496    &     -23.07	   &   23  &   --  &  --  \\
       A1795  &    0.0625     &      883    &     45.05    &      -25.57    &     1.978    &     -23.56	   &   27  &   --  &  --  \\
       A1831  &    0.0615     &      565    &     44.28    &      -25.63    &     1.266    &     -22.93	   &   21  &   --  &  --  \\
       A1983  &    0.0436     &      563    &     43.67    &      -24.87    &     1.272    &     -22.08	   &   14  &   --  &  --  \\
       A1991  &    0.0587     &      557    &     44.13    &      -25.51    &     1.250    &     -23.23	   &   20  &   --  &  --  \\
       A2107  &    0.0411     &      634    &     44.04    &      -24.90    &     1.435    &     -23.28	   &   27  &   --  &  --  \\
       A2124  &    0.0656     &      885    &     44.13    &      -25.51    &     1.980    &     -23.53	   &   30  &   38  &  19  \\
       A2149  &    0.0650     &      393    &     43.92    &      -25.61    &     0.879    &     -23.24	   &   --  &   20  &  --  \\
       A2169  &    0.0586     &      529    &     43.65    &      -24.80    &     1.188    &     -22.49	   &   --  &   10  &  --  \\
       A2256  &    0.0581     &     1353    &     44.85    &      -26.37    &     3.038    &     -23.40	   &   33  &   --  &  --  \\
       A2382  &    0.0618     &      998    &     43.96    &      -25.75    &     2.234    &     -22.84	   &   20  &   --  &  --  \\
       A2399  &    0.0579     &      781    &        44    &      -25.32    &     1.754    &     -22.60	   &   24  &   25  &  15  \\
      A2572a  &    0.0403     &      650    &     44.01    &      -24.94    &     1.472    &     -23.26	   &   10  &   --  &  --  \\
       A2589  &    0.0414     &      972    &     44.27    &      -24.78    &     2.200    &     -23.45	   &   22  &   --  &  --  \\
       A2593  &    0.0413     &      729    &     44.06    &      -24.97    &     1.650    &     -22.84	   &   --  &   23  &  --  \\
       A2657  &    0.0402     &      673    &      44.2    &      -24.85    &     1.524    &     -22.68	   &   21  &   --  &  --  \\
       A2734  &    0.0625     &      804    &     44.41    &      -25.06    &     1.802    &     -23.48	   &   28  &   --  &  --  \\
       A3128  &    0.0599     &      976    &     44.33    &      -26.26    &     2.190    &     -23.33	   &   47  &   --  &  --  \\
       A3158  &    0.0597     &     1117    &     44.73    &      -26.13    &     2.507    &     -23.82	   &   41  &   --  &  --  \\
       A3266  &    0.0589     &     1465    &     44.79    &      -26.28    &     3.288    &     -23.89	   &   40  &   --  &  --  \\
       A3376  &    0.0456     &      902    &     44.39    &      -25.04    &     2.037    &     -23.12	   &   20  &   --  &  --  \\
       A3395  &    0.0506     &     1195    &     44.45    &      -25.97    &     2.692    &     -23.39	   &   34  &   --  &  --  \\
       A3497  &    0.0677     &      787    &     44.16    &      -25.53    &     1.759    &     -22.45	   &   16  &   --  &  --  \\
      A3528a  &    0.0535     &     1093    &     44.12    &      -25.71    &     2.459    &     -23.77	   &   23  &   --  &  --  \\
      A3528b  &    0.0535     &      979    &      44.3    &      -25.74    &     2.203    &     -23.61	   &   14  &   --  &  --  \\
       A3530  &    0.0537     &      685    &     43.94    &      -25.55    &     1.541    &     -23.73	   &   26  &   --  &  --  \\
       A3532  &    0.0554     &      750    &     44.45    &      -25.91    &     1.686    &     -23.70	   &   37  &   --  &  --  \\
       A3556  &    0.0479     &      644    &     43.97    &      -25.58    &     1.453    &     -23.34	   &   25  &   --  &  --  \\
       A3558  &    0.0480     &      989    &      44.8    &      -26.38    &     2.232    &     -24.18	   &   52  &   --  &  --  \\
       A3560  &    0.0489     &      844    &     44.12    &      -25.68    &     1.903    &     -22.09	   &   19  &   --  &  --  \\
       A3667  &    0.0556     &     1170    &     44.94    &      -26.15    &     2.631    &     -23.97	   &   54  &   --  &  --  \\
       A3716  &    0.0462     &      855    &        44    &      -25.95    &     1.932    &     -22.94	   &   37  &   --  &  --  \\
       A3809  &    0.0620     &      631    &     44.35    &      -25.35    &     1.414    &     -22.85	   &   27  &   --  &  --  \\
       A3880  &    0.0584     &      893    &     44.27    &      -25.02    &     2.005    &     -23.07	   &   16  &   --  &  --  \\
       A4059  &    0.0475     &      843    &     44.49    &      -25.25    &     1.901    &     -23.64	   &   29  &   --  &  --  \\
     IIZW108  &    0.0493     &      579    &     44.34    &      -25.41    &     1.306    &     -23.77	   &   30  &   --  &  --  \\
       MKW3s  &    0.0450     &      575    &     44.43    &      -24.69    &     1.299    &     -22.72	   &   22  &   --  &  --  \\
      RX1022  &    0.0534     &      777    &     43.54    &      -25.16    &     1.748    &     -22.66	   &   --  &   11  &  --  \\
      RX1740  &    0.0430     &      596    &      43.7    &      -24.27    &     1.347    &     -22.41	   &   11  &   --  &  --  \\
       Z2844  &    0.0500     &      559    &     43.76    &      -23.93    &     1.260    &     -23.31	   &   21  &   --  &  --  \\
       Z8338  &    0.0473     &      747    &      43.9    &      -25.06    &     1.684    &     -23.15	   &   12  &   --  &  --  \\
       Z8852  &    0.0400     &      795    &     43.97    &      -25.30    &     1.800    &     -23.41	   &   18  &   --  &  --  \\
\enddata
\end{deluxetable}

\begin{table}
\begin{center}
\caption{FP coeff. for different galaxy samples, fitting algorithms and surface photometries.\label{tbl2}}
\medskip
\begin{tabular}{lccccccccc}
\tableline\tableline
Sample & a & b & c & $rms_a$  & $rms_b$ & $rms_c$ & $N_g$ & Fitting & Phot.\\
\tableline
\tableline
W+N+S & 1.152 & 0.320 & $-$8.56 & 0.021 & 0.004 & 0.095 & 1550 & MIST & WINGS \\
W+N+S & 1.293 & 0.322 & $-$8.91 & 0.021 & 0.003 & 0.002 & 1550 & ORTH & WINGS \\
\tableline
W+N & 1.113 & 0.319 & $-$8.45 & 0.021 & 0.004 & 0.102 & 1368 & MIST & WINGS \\
W+N & 1.258 & 0.329 & $-$8.99 & 0.022 & 0.003 & 0.003 & 1368 & ORTH & WINGS \\
\tableline
W+S & 1.332 & 0.318 & $-$8.93 & 0.050 & 0.008 & 0.198 & 282 & MIST & WINGS \\
W+S & 1.306 & 0.303 & $-$8.56 & 0.048 & 0.008 & 0.006 & 282 & ORTH & WINGS \\
\tableline
W+S & 1.297 & 0.319 & $-$8.87 & 0.050 & 0.008 & 0.198 & 282 & MIST & SDSS \\
\tableline
COMA & 1.239 & 0.342 & $-$9.15 & 0.080 & 0.013 & 0.310 & 80 & MIST & JORG \\
COMA & 1.439 & 0.345 & $-$9.67 & 0.077 & 0.013 & 0.010 & 80 & ORTH & JORG \\
\tableline
\end{tabular}
\end{center}
\end{table}

\begin{deluxetable}{rccccccc|ccccccc}
\rotate
\tabletypesize{\small}
\tablewidth{0pt}
\tablecaption{FP MIST coefficients of the individual clusters.\label{tbl3}}
\tablehead{
& \multicolumn{7}{c}{All Galaxies} & \multicolumn{7}{c}{Galaxies with $M_V<$-19.5} \\
\colhead{Cluster} & \colhead{$N_g$} & \colhead{a} & \colhead{rms(a)} & \colhead{b} & \colhead{rms(b)} & \colhead{c} & \colhead{rms(c)} & \colhead{$N_g$} & \colhead{a} & \colhead{rms(a)} & \colhead{b} & \colhead{rms(b)} & \colhead{c} & \colhead{rms(c)}}
\startdata
A0085  & 63 & 1.137 & 0.083 & 0.304 & 0.013 &  -8.24 & 0.33 &  52 & 1.013 & 0.076 & 0.289 & 0.014 &  -7.64 & 0.36  \\ 	    
A1069  & 20 & 1.236 & 0.216 & 0.275 & 0.030 &  -7.81 & 0.75 &  20 & 1.236 & 0.216 & 0.275 & 0.030 &  -7.81 & 0.75  \\ 	    
A119   & 48 & 1.289 & 0.127 & 0.289 & 0.018 &  -8.21 & 0.54 &  45 & 1.169 & 0.113 & 0.287 & 0.018 &  -7.90 & 0.52  \\ 	    
A1291  & 13 & 1.415 & 0.222 & 0.381 & 0.016 & -10.37 & 0.62 &  11 & 1.635 & 0.389 & 0.377 & 0.018 & -10.78 & 1.09  \\ 	    
A133   & 23 & 1.462 & 0.161 & 0.371 & 0.009 & -10.27 & 0.41 &  23 & 1.462 & 0.161 & 0.371 & 0.009 & -10.27 & 0.41  \\ 	    
A160   & 23 & 1.580 & 0.247 & 0.350 & 0.040 & -10.10 & 0.89 &  19 & 1.743 & 0.303 & 0.363 & 0.050 & -10.73 & 1.22  \\ 	    
A1631a & 22 & 1.214 & 0.093 & 0.289 & 0.014 &  -8.06 & 0.23 &  21 & 1.187 & 0.106 & 0.290 & 0.014 &  -8.02 & 0.23  \\ 	    
A1644  & 41 & 1.030 & 0.114 & 0.323 & 0.019 &  -8.35 & 0.56 &  40 & 1.088 & 0.124 & 0.329 & 0.020 &  -8.61 & 0.60  \\ 	    
A1668  & 23 & 0.781 & 0.192 & 0.274 & 0.030 &  -6.80 & 0.90 &  23 & 0.781 & 0.192 & 0.274 & 0.030 &  -6.80 & 0.90  \\ 	    
A168   & 12 & 1.279 & 0.129 & 0.344 & 0.058 &  -9.27 & 1.16 &  11 & 1.111 & 0.092 & 0.316 & 0.051 &  -8.32 & 1.03  \\ 	
A1795  & 27 & 0.774 & 0.120 & 0.255 & 0.029 &  -6.41 & 0.79 &  27 & 0.774 & 0.120 & 0.255 & 0.029 &  -6.41 & 0.79  \\ 	    
A1831  & 21 & 0.725 & 0.116 & 0.325 & 0.020 &  -7.74 & 0.38 &  21 & 0.725 & 0.116 & 0.325 & 0.020 &  -7.74 & 0.38  \\ 	    
A1983  & 14 & 1.044 & 0.156 & 0.271 & 0.039 &  -7.33 & 0.83 &  13 & 0.995 & 0.160 & 0.257 & 0.040 &  -6.93 & 0.85  \\ 	    
A1991  & 20 & 0.933 & 0.183 & 0.253 & 0.051 &  -6.68 & 1.19 &  20 & 0.933 & 0.183 & 0.253 & 0.051 &  -6.68 & 1.19  \\ 	    
A2107  & 27 & 0.993 & 0.127 & 0.269 & 0.028 &  -7.22 & 0.64 &  25 & 0.981 & 0.151 & 0.296 & 0.014 &  -7.72 & 0.46  \\ 	    
A2124  & 49 & 1.065 & 0.131 & 0.317 & 0.014 &  -8.27 & 0.50 &  48 & 0.992 & 0.114 & 0.314 & 0.014 &  -8.05 & 0.44  \\ 	    
A2149  & 20 & 1.159 & 0.127 & 0.321 & 0.023 &  -8.58 & 0.62 &  20 & 1.159 & 0.127 & 0.321 & 0.023 &  -8.58 & 0.63  \\ 	    
A2169  & 10 & 1.500 & 0.165 & 0.331 & 0.047 &  -9.55 & 0.78 &   8 & 1.441 & 0.136 & 0.286 & 0.053 &  -8.48 & 0.98  \\ 	    
A2256  & 33 & 0.885 & 0.129 & 0.302 & 0.018 &  -7.55 & 0.43 &  33 & 0.885 & 0.129 & 0.302 & 0.018 &  -7.55 & 0.44  \\ 	    
A2382  & 20 & 1.547 & 0.199 & 0.314 & 0.019 &  -9.32 & 0.54 &  20 & 1.547 & 0.199 & 0.314 & 0.019 &  -9.32 & 0.54  \\ 	    
A2399  & 34 & 1.154 & 0.113 & 0.338 & 0.017 &  -8.91 & 0.39 &  30 & 1.002 & 0.092 & 0.324 & 0.014 &  -8.28 & 0.33  \\ 	    
A2572a & 10 & 1.157 & 0.267 & 0.299 & 0.047 &  -8.20 & 1.45 &  10 & 1.157 & 0.267 & 0.299 & 0.047 &  -8.20 & 1.45  \\ 	    
A2589  & 22 & 1.016 & 0.188 & 0.315 & 0.038 &  -8.20 & 1.06 &  20 & 0.854 & 0.163 & 0.303 & 0.039 &  -7.59 & 1.07  \\ 	    
A2593  & 23 & 1.559 & 0.221 & 0.320 & 0.052 &  -9.48 & 1.28 &  15 & 0.698 & 0.089 & 0.189 & 0.028 &  -4.91 & 0.74  \\ 	    
A2657  & 21 & 1.059 & 0.165 & 0.336 & 0.031 &  -8.70 & 0.84 &  21 & 1.059 & 0.165 & 0.336 & 0.031 &  -8.70 & 0.84  \\ 	    
A2734  & 28 & 1.071 & 0.198 & 0.325 & 0.028 &  -8.52 & 0.93 &  28 & 1.071 & 0.198 & 0.325 & 0.028 &  -8.52 & 0.93  \\ 	    
A3128  & 47 & 1.329 & 0.133 & 0.365 & 0.024 &  -9.86 & 0.63 &  47 & 1.329 & 0.133 & 0.365 & 0.024 &  -9.86 & 0.63  \\ 	    
A3158  & 41 & 1.189 & 0.090 & 0.306 & 0.019 &  -8.34 & 0.40 &  41 & 1.189 & 0.090 & 0.306 & 0.019 &  -8.34 & 0.40  \\ 	    
A3266  & 40 & 0.976 & 0.105 & 0.337 & 0.015 &  -8.51 & 0.40 &  40 & 0.976 & 0.105 & 0.337 & 0.015 &  -8.51 & 0.40  \\ 	    
A3376  & 20 & 1.174 & 0.210 & 0.293 & 0.032 &  -8.07 & 1.01 &  20 & 1.174 & 0.210 & 0.293 & 0.032 &  -8.07 & 1.01  \\ 	    
A3395  & 34 & 1.066 & 0.091 & 0.374 & 0.021 &  -9.47 & 0.44 &  34 & 1.066 & 0.091 & 0.374 & 0.021 &  -9.47 & 0.44  \\ 	    
A3497  & 16 & 0.731 & 0.151 & 0.274 & 0.018 &  -6.64 & 0.59 &  16 & 0.731 & 0.151 & 0.274 & 0.018 &  -6.64 & 0.59  \\ 	    
A3528a & 23 & 0.730 & 0.150 & 0.334 & 0.019 &  -7.90 & 0.61 &  23 & 0.730 & 0.150 & 0.334 & 0.019 &  -7.90 & 0.61  \\ 	    
A3528b & 14 & 1.246 & 0.163 & 0.256 & 0.021 &  -7.53 & 0.64 &  13 & 1.236 & 0.197 & 0.256 & 0.025 &  -7.49 & 0.84  \\ 	    
A3530  & 26 & 0.956 & 0.107 & 0.313 & 0.017 &  -8.03 & 0.35 &  26 & 0.956 & 0.107 & 0.313 & 0.017 &  -8.03 & 0.35  \\ 	    
A3532  & 37 & 1.101 & 0.099 & 0.326 & 0.020 &  -8.60 & 0.42 &  37 & 1.101 & 0.099 & 0.326 & 0.020 &  -8.60 & 0.42  \\ 	    
A3556  & 25 & 1.224 & 0.171 & 0.384 & 0.028 & -10.08 & 0.76 &  24 & 1.155 & 0.177 & 0.381 & 0.027 &  -9.85 & 0.78  \\ 	    
A3558  & 52 & 1.014 & 0.079 & 0.360 & 0.016 &  -9.06 & 0.38 &  52 & 1.014 & 0.079 & 0.360 & 0.016 &  -9.06 & 0.38  \\ 	    
A3560  & 19 & 1.284 & 0.225 & 0.303 & 0.046 &  -8.52 & 0.78 &  19 & 1.284 & 0.225 & 0.303 & 0.046 &  -8.52 & 0.78  \\ 	    
A3667  & 54 & 1.208 & 0.089 & 0.326 & 0.019 &  -8.82 & 0.47 &  54 & 1.208 & 0.089 & 0.326 & 0.019 &  -8.82 & 0.48  \\ 	    
A3716  & 37 & 1.277 & 0.166 & 0.323 & 0.019 &  -8.85 & 0.48 &  37 & 1.277 & 0.166 & 0.323 & 0.019 &  -8.85 & 0.48  \\ 	    
A376   & 27 & 1.067 & 0.108 & 0.307 & 0.019 &  -8.03 & 0.51 &  27 & 1.067 & 0.108 & 0.307 & 0.019 &  -8.03 & 0.51  \\ 	
A3809  & 27 & 0.903 & 0.073 & 0.329 & 0.017 &  -8.18 & 0.33 &  27 & 0.903 & 0.073 & 0.329 & 0.017 &  -8.18 & 0.33  \\ 	    
A3880  & 16 & 1.096 & 0.115 & 0.397 & 0.057 &  -9.96 & 1.22 &  16 & 1.096 & 0.115 & 0.397 & 0.057 &  -9.96 & 1.22  \\ 	    
A4059  & 29 & 1.149 & 0.134 & 0.336 & 0.020 &  -8.91 & 0.58 &  26 & 1.127 & 0.137 & 0.343 & 0.024 &  -9.00 & 0.66  \\ 	    
A548b  & 22 & 0.991 & 0.120 & 0.325 & 0.015 &  -8.36 & 0.52 &  18 & 0.941 & 0.082 & 0.317 & 0.010 &  -8.09 & 0.32  \\ 	    
A602   & 18 & 1.180 & 0.231 & 0.361 & 0.048 &  -9.40 & 1.37 &  16 & 0.948 & 0.175 & 0.330 & 0.038 &  -8.25 & 1.02  \\ 	
A671   & 16 & 1.123 & 0.161 & 0.273 & 0.020 &  -7.58 & 0.64 &  14 & 1.175 & 0.218 & 0.296 & 0.028 &  -8.16 & 0.96  \\ 	
A754   & 46 & 0.993 & 0.107 & 0.317 & 0.016 &  -8.17 & 0.40 &  46 & 0.993 & 0.107 & 0.317 & 0.016 &  -8.17 & 0.40  \\ 	
A780   & 16 & 1.364 & 0.215 & 0.325 & 0.032 &  -9.14 & 0.96 &  16 & 1.364 & 0.215 & 0.325 & 0.032 &  -9.14 & 0.96  \\ 	
A957x  & 29 & 1.266 & 0.113 & 0.319 & 0.014 &  -8.80 & 0.39 &  23 & 1.085 & 0.093 & 0.312 & 0.012 &  -8.24 & 0.31  \\ 	    
A970   & 25 & 1.156 & 0.223 & 0.317 & 0.022 &  -8.49 & 0.73 &  25 & 1.156 & 0.223 & 0.317 & 0.022 &  -8.49 & 0.72  \\ 	
IIZW108& 30 & 0.957 & 0.137 & 0.244 & 0.031 &  -6.64 & 0.79 &  29 & 0.932 & 0.138 & 0.243 & 0.031 &  -6.57 & 0.78  \\ 	    
MKW3s  & 22 & 1.112 & 0.151 & 0.260 & 0.024 &  -7.31 & 0.59 &  20 & 1.099 & 0.132 & 0.259 & 0.028 &  -7.24 & 0.62  \\ 	    
RX1022 & 11 & 1.230 & 0.146 & 0.305 & 0.030 &  -8.47 & 0.30 &   9 & 1.138 & 0.147 & 0.291 & 0.056 &  -7.97 & 1.04  \\ 	    
RX1740 & 11 & 1.327 & 0.408 & 0.427 & 0.047 & -11.08 & 1.41 &  11 & 1.327 & 0.408 & 0.427 & 0.047 & -11.08 & 1.41  \\ 	    
Z2844  & 21 & 0.961 & 0.192 & 0.213 & 0.021 &  -6.00 & 0.54 &  18 & 0.935 & 0.201 & 0.226 & 0.022 &  -6.21 & 0.53  \\ 	    
Z8338  & 12 & 0.933 & 0.244 & 0.273 & 0.047 &  -7.12 & 1.27 &  10 & 0.685 & 0.194 & 0.230 & 0.033 &  -5.69 & 0.98  \\ 	    
Z8852  & 18 & 0.761 & 0.097 & 0.339 & 0.023 &  -8.13 & 0.52 &  17 & 0.731 & 0.115 & 0.337 & 0.023 &  -8.02 & 0.54  \\     
\enddata
\end{deluxetable}

\begin{table}
\begin{center}
\caption{Statistics of the measured coefficients for the MIST fits of the FP.\label{tbl4}}
\medskip
\begin{tabular}{cccccl}
\tableline\tableline
Sample & Coefficient & Average & St. Dev. & Median & Notes \\
\tableline
W+N+S& a & 1.121$\pm$0.027 & 0.207 & 1.123 & (all clusters) \\
     & b & 0.316$\pm$0.005 & 0.040 & 0.319 & \\
     & c &$-$8.41$\pm$0.137& 1.052 &$-$8.35& \\
\tableline
W+N+S& a & 1.108$\pm$0.037 & 0.207 & 1.071 & (clusters with $N_g>N_{med}$) \\
     & b & 0.321$\pm$0.006 & 0.033 & 0.323 & \\
     & c &$-$8.49$\pm$0.172& 0.956 &$-$8.49& \\
\tableline
W+N  & a & 1.081$\pm$0.029 & 0.211 & 1.066 & (all clusters) \\
     & b & 0.311$\pm$0.006 & 0.041 & 0.315 & \\
     & c &$-$8.23$\pm$0.140& 1.012 &$-$8.20& \\
\tableline
W+N  & a & 1.047$\pm$0.033 & 0.176 & 1.064 & (clusters with $N_g>N_{med}$) \\
     & b & 0.319$\pm$0.006 & 0.034 & 0.323 & \\
     & c &$-$8.31$\pm$0.180& 0.953 &$-$8.34& \\
\tableline
W+S   & a & 1.308$\pm$0.052 & 0.195 & 1.279 & (all clusters) \\
      & b & 0.327$\pm$0.008 & 0.030 & 0.324 & \\
      & c &$-$9.04$\pm$0.252& 0.943 &$-$8.73& \\
\tableline
W+S   & a & 1.226$\pm$0.115 & 0.230 & 1.201 & (clusters with $N_g>N_{med}$) \\
      & b & 0.313$\pm$0.005 & 0.011 & 0.320 & \\
      & c &$-$8.59$\pm$0.335& 0.671 &$-$8.08& \\
\tableline
\end{tabular}
\end{center}
\end{table}


\end{document}